\documentclass[sigconf]{acmart}

\copyrightyear{2024}
\acmYear{2024}
\setcopyright{acmlicensed}
\acmConference[WSDM '24] {Proceedings of the 17th ACM International Conference on Web Search and Data Mining}{March 4--8, 2024}{Merida, Mexico.}
\acmBooktitle{Proceedings of the 17th ACM International Conference on Web Search and Data Mining (WSDM '24), March 4--8, 2024, Merida, Mexico}
\acmPrice{15.00}
\acmISBN{979-8-4007-0371-3/24/03}
\acmDOI{10.1145/XXXXXX.XXXXXX}

\usepackage{multirow} 
\usepackage{subfigure} 
\usepackage{enumitem}
\usepackage{algorithm}
\usepackage{algorithmic}
\AtBeginDocument{%
  }

\begin{document}


\title{Intent Contrastive Learning with Cross Subsequences for Sequential Recommendation}
\author{Xiuyuan Qin}
\authornote{These authors contributed equally to this work.}
\affiliation{
  \city{Soochow University}
  \country{China}}
\email{20215227016@stu.suda.edu.cn}

\author{Huanhuan Yuan}
\authornotemark[1]
\affiliation{
  \city{Soochow University}
  \country{China}}
\email{hhyuan@stu.suda.edu.cn}

\author{Pengpeng Zhao}
\authornote{Corresponding author.}
\affiliation{
  \city{Soochow University}
  \country{China}}
\email{ppzhao@suda.edu.cn}

\author{Guanfeng Liu}
\affiliation{
  \city{Macquarie University}
  \country{Australia}}
\email{guanfeng.liu@mq.edu.au}

\author{Fuzhen Zhuang}
\affiliation{
  \institution{Institute of Artificial Intelligence \& SKLSDE, School of Computer Science}
  \city{Beihang University}
  \country{China}}
\email{zhuangfuzhen@buaa.edu.cn}

\author{Victor Sheng}
\affiliation{
  \city{Texas Tech University}
  \country{United States}}
\email{victor.sheng@ttu.edu}








\renewcommand{\shortauthors}{Xiuyuan Qin et al.}

\begin{abstract}
The user purchase behaviors are mainly influenced by their intentions (e.g., buying clothes for decoration, buying brushes for painting, etc.). 
Modeling a user's latent intention can significantly improve the performance of recommendations. 
Previous works model users' intentions by considering the predefined label in auxiliary information or introducing stochastic data augmentation to learn purposes in the latent space.  
However, the auxiliary information is sparse and not always available for recommender systems, and introducing stochastic data augmentation may introduce noise and thus change the intentions hidden in the sequence. 
Therefore, leveraging user intentions for sequential recommendation (SR) can be challenging because they are frequently varied and unobserved. 
In this paper, Intent contrastive learning with Cross Subsequences for sequential Recommendation (ICSRec) is proposed to model users' latent intentions. 
Specifically, ICSRec first segments a user's sequential behaviors into multiple subsequences by using a dynamic sliding operation and takes these subsequences into the encoder to generate the representations for the user's intentions. 
To tackle the problem of no explicit labels for purposes, ICSRec assumes different subsequences with the same target item may represent the same intention and proposes a coarse-grain intent contrastive learning to push these subsequences closer. 
Then, fine-grain intent contrastive learning is mentioned to capture the fine-grain intentions of subsequences in sequential behaviors.
Extensive experiments conducted on four real-world datasets demonstrate the superior performance of the proposed ICSRec\footnote{Our code is available at \url{https://github.com/QinHsiu/ICSRec}.} model compared with baseline methods. 
\end{abstract}




\maketitle

\section{Introduction}
\label{sec:0}
Recommender systems are widely studied in academia and industry for their ability to efficiently alleviate data overload by accurately capturing users' preferences and providing personalized recommendation suggestions~\cite{Rec_survey,Rec_survey01,Rec_survey02}. 
Sequential Recommendation (SR)~\cite{SASRec,BERT4Rec,STOSA,SRS,SSL}, one of the best recommendation models for dynamically modeling user preferences, focus on predicting the next item that users are likely to interact with based on their chronological behaviors. 
\begin{figure}[t]
\centerline{\includegraphics[width=0.5\textwidth]{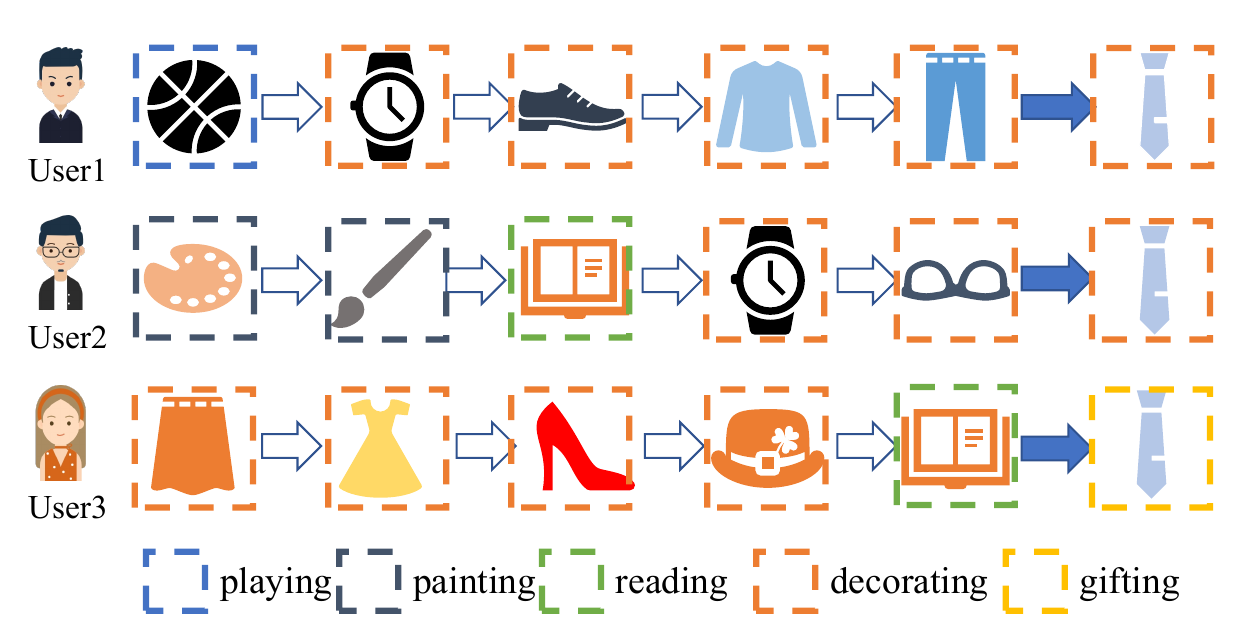}}\caption{An example showing different interaction sequences exist with subsequences of the same target item.} 
\label{fig1}
\end{figure}

Users' purchase behaviors mainly depend on their intentions (e.g., buying a watch for decoration, buying books for reading, etc.)~\cite{ICLRec,IOCRec}. 
Leveraging such intentions for SR can significantly improve the performance and robustness~\cite{CAFE}, and this has attracted widespread attention. 
The methods for modeling intentions can be divided into two categories based on whether or not to use auxiliary information (e.g., user action types, item category information, etc.). 
Some exciting works~\cite{ASLI,CoCoRec,CAFE} model users' intentions by constructing an auxiliary task (e.g., predicting the next item's category, predicting the user's next action type, etc.).  
Since such information is more sparse, does not always accurately reflect the user's intentions, and may not be readily accessible~\cite{ICLRec}, some other exciting works choose to capture user intentions in the latent space. 
DSSRec~\cite{DSSRec} proposes a seq2seq training strategy, which infers the intentions based on individual sequence representation via clustering. 
SINE~\cite{SINE} proposes a sparse interest extraction module to adaptively infer the interacted intentions of a user from a large pool of intention groups. 
ICLRec~\cite{ICLRec} obtains the intent prototype representation from the embedded space of all user behavior sequences via clustering and builds a contrastive learning task to leverage the learned intentions into the SR model. 
IOCRec~\cite{IOCRec} extracts intent representations from two stochastic augmented views for contrastive learning and further improves the performance by alleviating the noise problem. 

While the above attempts to construct contrastive learning tasks to learn the user's intentions, they usually model each user interaction sequence individually and thus ignore the correlation between users with similar subsequence patterns~\cite{GCL4SR}. 
Considering the example illustrated in Figure~\ref{fig1}, User1 and User2 may purchase the item `watch' for decorating at different steps. 
Meanwhile, User2 and User3 may buy the item `book' for learning at different steps. 
Different users may have the same intention to purchase the same item at different moments. 
However, modeling sequences as a whole does not fully utilize the intent supervisory signals hidden in the history of different user interactions. 
Furthermore, the intention to buy the same item in different contexts may be different~\cite{challenge3}. 
As shown in Figure~\ref{fig1}, all three Users buy the item `tie'. 
User3 may buy it as a gift, but User1 and User2 may buy it for decoration. 
Ignoring this restriction may result in the model's failure to learn the representation of the user's intention, thus decreasing the performance. 

To address the above issues, we propose a novel model named ICSRec (i.e., Intent contrastive learning with Cross subsequences for Sequential Recommendation), which utilizes cross subsequences patterns to construct intent supervisory signals for intent representation learning. 
Specially, we first process the original training sequence into several subsequences via a split operation, and we put subsequences with the same target item into the same set to construct coarse-grain intent supervisory signals. 
ICSRec assumes that two different subsequences with the same target item have the same intention. 
Therefore, a coarse-grain intent contrastive learning method is introduced to put the two subsequences with the same intention closer. 
To alleviate the problem that the same item may represent different intentions in different contexts, we further propose a fine-grain intent contrastive leaning to make the subsequence closer to its intention prototype obtained by clustering. 
Finally, ICSRec achieves a significant improvement (average increase of 24\%) compared with the State-Of-The-Art baselines on all datasets. 

We summarize the contributions of this work below:
\begin{itemize}[leftmargin=*]
\item ICSRec first utilizes a similar pattern hidden in cross subsequences for intent representation learning in the SR task. 
\item We propose two modules (i.e., coarse-grain intent learning module and fine-grain intent learning module ) to model the user's intention from different dimensions. 
\item Extensive experiments demonstrate that ICSRec achieves state-of-the-art performances on four publicity datasets.
\end{itemize}

\begin{figure*}[t]
     \centering
          \includegraphics[width=0.8\linewidth]{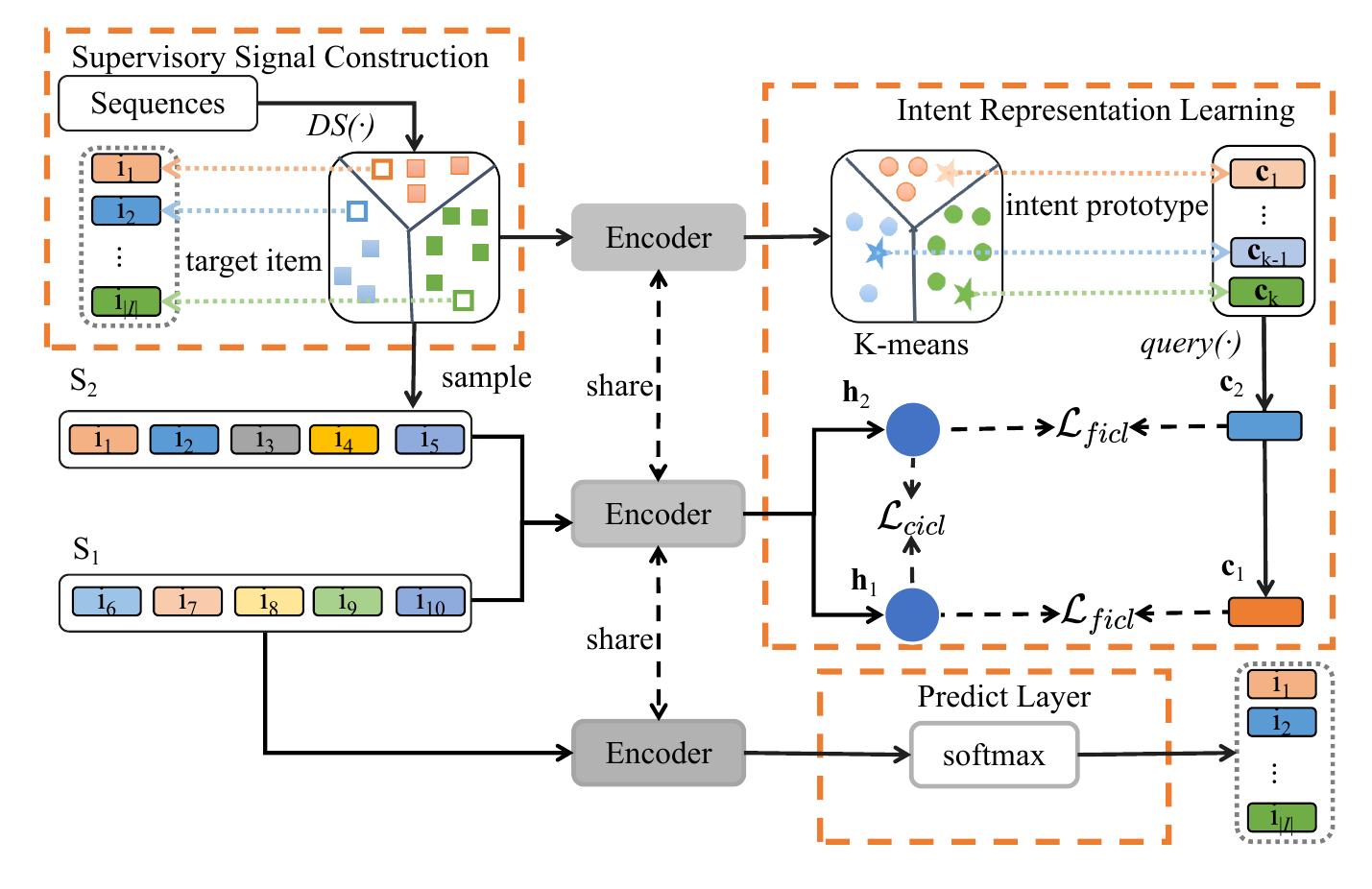}
     \caption{The model architecture of ICSRec. 
     Where $S_{2}$ and $S_{1}$ denote two subsequences with the same target item. 
     $\mathbf{h}_{2}$ and $\mathbf{h}_{1}$ denote two coarse-grain intentions obtained by encoder $f_{\theta}(\cdot)$. 
     $\mathbf{c}_{2}$ and $\mathbf{c}_{1}$ denote two fine-grain intentions obtained via clustering. 
     }
     \label{model}
\end{figure*}

\section{The Proposed Model}
\subsection{Problem Definition}
Sequential Recommendation (SR) is to recommend the next item that the user will interact with based on his/her historical interaction data. 
Assuming that $U$ and $I$ are the sets of users and items, $u\in U$ has a sequence of interacted items $S^u=\{s_1^u, s_2^u,..., s_{|S^u|}^u\}$, and $s^{u}_{t} \in \mathcal{I} (1 \leq t \leq |S^{u}|)$ represents an interacted item at position $t$ of user $u$ within the sequence, where $|S^{u}|$ denotes the sequence length. 
Given the historical interactions $S^{u}$, the goal of SR is to recommend an item from the set of items $\mathcal{I}$ that the user $u$ may interact with at the $|S^{u}|+1$ step, which can be formulated as follows:
 \begin{equation}
 \arg\max_{i \in \mathcal{I}} P(s^{u}_{|S^{u}|+1}=i|S^{u}).
     \label{eq1}
 \end{equation}
\subsection{Overall Framework}
Figure~\ref{model} shows the overall framework of ICSRec. 
It mainly contains three parts, 1) Supervisory Signal Construction, 2) Intent Representation Learning, and 3) Predict Layer. 
In the first part, we first segment all training sequences into multiple subsequences by the $DS(\cdot)$ operation via Eq.(\ref{eq2}). 
Then, we put all subsequences with the same target item into a set to construct coarse-grain intent supervisory signals. 
Next, we set up two auxiliary tasks for intent representation learning: Coarse-grain Intent Contrastive Learning (CICL) and Fine-grain Intent Contrastive Learning (FICL). 
In the CICL task, for each subsequence, we randomly sample another subsequence from the set as its positive sample with the same target item. 
Then, we directly make the coarse-grain intent representation of the two subsequences in the latent space closer. 
In the FICL task, we first cluster the coarse-grain intent representations obtained from all subsequences and then make the coarse-grain intent representation closer to its fine-grain intent prototype. 
In addition, we utilize the learned intent representation to predict the next item that the user may interact with. 
Finally, we jointly optimize these three objectives.   
\subsection{Supervisory Signal Construction}
To investigate the same subsequence patterns across user interaction sequences, we first split the original sequence into multiple subsequences via $DS(\cdot)$ operation~\cite{DataAug}. Specially, given a sequence $S^u=\{s_1^u,s_2^u,..., s_{|S^u|}^{u}\}$, the operation can be formulated as follows:
\begin{eqnarray}
		DS(S^{u})=
		\begin{cases}
		\{\{s^{u}_{1},i^{u}_{2}\},\cdots, \{s^{u}_{1},\\ \cdots,s^{u}_{|S^{u}|-1},i^{u}_{|S^{u}|}\}\} & |S^{u}|\leq n+1 \\ \\
			\{ DS(S^{u}_{n+1}),\cdots,\\ \{s^{u}_{|S^{u}|-n}, \cdots,i^{u}_{|S^{u}|}\}\} & |S^{u}| > n+1.
		\end{cases}
		\label{eq2}
\end{eqnarray}
where $n$ represents the max sequence length and $i^{u}_{k}(1\leq k \leq |S^{u}|)$ represents the target item of the subsequence. 
After the segmentation operation on all training sequences, we put all obtained subsequences into different sets $T=\{T_{1}, T_{2},..., T_{|I|}\}$, where $T_{a}$ represents the subsequence set with the target item $i_{a}\in I(1\leq a\leq |I|)$, to construct the coarse-grain intent supervisory signals. 
\subsection{Encoder}
Firstly, the whole item sets $\mathcal{I}$ are embedded into the same space~\cite{SASRec,BERT4Rec} and generate the item embedding matrix $\mathbf{M} \in \mathbb{R}^{|\mathcal{I}| \times d}$. 
Given the input sequence $S^{u}$, the embedding of the sequence $S^{u}$ is initialized to $\mathbf{e}^{u} \in \mathbb{R}^{n \times d}$, which can formulated as follows:
\begin{equation}
    \mathbf{e}^{u}=\mathbf{m}^{u}+\mathbf{p}^{u},
    \label{eq3}
\end{equation}
where $\mathbf{m}^{u} \in \mathbb{R}^{n\times d}$, represents the item's embedding, 
$\mathbf{p}^{u} \in \mathbb{R}^{n\times d}$ represents the position embedding and $n$ represents the length of the sequence, respectively. 

Due to its superiority in modeling sequential tasks, we choose SASRec~\cite{SASRec} as the backbone model, represented as $f_{\theta}(\cdot)$, which uses the Transformer~\cite{Att} to learn the representation of the sequence. 
Given the sequence embedding $\mathbf{e}^{u}$, the output representation $\mathbf{H}^{u} \in \mathbb{R}^{n \times d}$ is calculated as:
\begin{equation}
\mathbf{H}^{u}=f_{\theta}(\mathbf{e}^{u}).
    \label{eq4}
\end{equation}
where $\theta$ represents the parameters of the sequential model. 
The last vector $\mathbf{h}^{u}_{n} \in \mathbb{R}^{d}$ in $\mathbf{H}^{u}=[\mathbf{h}^{u}_{1},...,\mathbf{h}^{u}_{n}]$ is chosen as the intent representation of the sequence. 
\subsection{Intent Representation Learning} 
\noindent\textbf{Coarse-grain Intent Contrastive Learning.} 
Since different subsequences with the same target item may represent the same intention. 
Thus, we directly bring the intent representations of two subsequences with the same target item closer together in the latent space via contrastive learning. 
Given a subsequence $S_{1}$ with the target item $i_{a}$, we first randomly sample a subsequence $S_{2}$ from $T_{a}$, and then calculate the coarse-grain intent representations, $\mathbf{h}_{1}$ and $\mathbf{h}_{2}$, of two subsequences by Eq.(\ref{eq4}), respectively. 
Then, we utilize contrast learning to bring two coarse-force intent representations closer to each other in the latent space. The contrastive loss can be formulated as follows:
\begin{equation}
    \mathcal{L}_{cicl}=\mathcal{L}_{con}(\mathbf{h}_{1},\mathbf{h}_{2}),
\label{eq5}
\end{equation}
and
\begin{equation}
\begin{split}
\mathcal{L}_{con}(\mathbf{x}^{1},\mathbf{x}^{2})=-\log\frac{e^{s(\mathbf{x}^{1}, \mathbf{x}^{2})/\tau}}{e^{s(\mathbf{x}^{1}, \mathbf{x}^{2})/\tau}+\underset{\mathbf{x} \notin \mathcal{F}}{\sum}e^{s(\mathbf{x}^{1},\mathbf{x})/\tau}}\\
-\log\frac{e^{s(\mathbf{x}^{2}, \mathbf{x}^{1})/\tau}}{e^{s(\mathbf{x}^{2}, \mathbf{x}^{1})/\tau}+\underset{\mathbf{x} \notin \mathcal{F}}{\sum}e^{s(\mathbf{x}^{2},\mathbf{x})/\tau}}.
\end{split}
    \label{eq6}
\end{equation}

where $(\mathbf{x}^{1}, \mathbf{x}^{2})$ represents a pair of positive sample's embedding, $s(\cdot)$ represents the inner product, $\tau$ is the temperature parameter, and $\mathcal{F}$ is a set of negative samples that have the same label with two positive pairs in the mini-batch. 
Specially, we do not use InfoNCE~\cite{simclr,moco} directly to calculate the contrastive loss, because treating other 2($|\mathbf{B}|$-1) views within the same batch as negative samples may introduce false negative problems (e.g., there have more than one pair of subsequences that have the same target item in a mini-batch.), where $|\mathbf{B}|$ denotes the batch size. 
So we propose a simple strategy called False Negative Mask (FNM) to mask the effects by not contrasting them~\cite{ICLRec}.

\noindent\textbf{Fine-grain Intent Contrastive Learning.} 
Since the intention to buy the same item may fall into different categories in different contexts. 
Thus, we assume that there are $K$ types of users' intentions and that the intention to purchase the same item in different contexts can potentially belong to different types. 
We first put all subsequences into the encoder to obtain all the coarse-grain intent representations via Eq.(~\ref{eq4}). 
Then we use all the obtained outputs for K-means clustering via faiss~\cite{faiss} and treat the center of the clusters as the fine-grain intent representation. 
The intention prototype we obtained is represented as $\mathbf{C}=\{\mathbf{c}^{k}\}_{k=1}^{K}$, where $\mathbf{c}^{k} \in \mathbb{R}^{d}$ represents the $k$-th intention prototype. 
Then, we obtain the fine-grain intent representation of the two subsequences by $query(\cdot)$ operation. 
Specially, we choose the nearest intention prototype $\mathbf{c}_{1}$ to $\mathbf{h}_{1}$ and $\mathbf{c}_{2}$ to $\mathbf{h}_{2}$, respectively, from the set $\mathbf{C}=\{\mathbf{c}^{k}\}_{k=1}^{K}$ as the fine-grain intent representation of $\mathbf{h}_{1}$ and $\mathbf{h}_{2}$, which is as follows:
\begin{equation}
\mathbf{c}_1,\mathbf{c}_2=query(\mathbf{h}_1,\mathbf{C}),query(\mathbf{h}_2,\mathbf{C}).
\label{eq7}
\end{equation}
To avoid introducing false negative problems (e.g., there have more than one pair of subsequences that have the same fine-grain intent representation in a mini-batch.). 
We use Eq.(~\ref{eq6}) to calculate the contrastive loss as follows:
\begin{equation}
 \mathcal{L}_{ficl}= \mathcal{L}_{con}(\mathbf{h}_{1},\mathbf{c}_{1})+\mathcal{L}_{con}(\mathbf{h}_{2},\mathbf{c}_{2}).
    \label{eq8}
\end{equation} 
\subsection{Prediction Layer}
In SR, the prediction of the next item can be viewed as a classification task based on the set of all items. 
Therefore, we use the learned intent representation to calculate the probability of the next item that the user may interact with. 
Given the learned intent representation $\mathbf{h}^{u}\in \mathbb{R}^{d}$ and the item embedding matrix $\mathbf{M}$, the Eq.(\ref{eq1}) is equivalent to minimizing the cross$\mbox{-}$entropy loss as follows: 
\begin{equation}
    \mathcal{L}_{Rec}=-1*\hat{\mathbf{y}}[g]+\log(\sum_i^I\exp(\hat{\mathbf{y}}[i]))),
    \label{eq9}
\end{equation}
and
\begin{equation}
    \hat{\mathbf{y}} = softmax(\mathbf{h}^{u}\mathbf{M}^{T}).
    \label{eq10}
\end{equation}
where $\hat{\mathbf{y}} \in \mathbb{R}^{|I|}$ represents the predictive score of all items, and $g \in I$ represents the ground$\mbox{-}$truth of the sequence. 
\subsection{Multi Task Learning}
We use a multi$\mbox{-}$task learning paradigm to jointly optimize the main sequential prediction task and the other two auxiliary learning objectives. 
In which Eq.(\ref{eq9}) is to optimize the main next item predict task, Eq.(\ref{eq5}) is to optimize the CICL task, and Eq.(\ref{eq8}) is to optimize the FICL task. 
The final training loss function is as follows:
\begin{equation}
\mathcal{L}=\mathcal{L}_{Rec}+\lambda\cdot{\mathcal{L}_{cicl}+\beta\cdot{\mathcal{L}_{ficl}}}.
    \label{eq11}
\end{equation}
where $\lambda$ and $\beta$ are hyper-parameters that need to be tuned.
\begin{table}[t]
  \centering
      \renewcommand{\arraystretch}{1.0}
    \resizebox{\linewidth}{!}{
    \begin{tabular}{cccccccccc}
    \hline
    \multicolumn{2}{l|}{Dataset} & \multicolumn{2}{c}{Sports} & \multicolumn{2}{c}{Beauty} & \multicolumn{2}{c}{Toys} & \multicolumn{2}{c}{ML-1M} \\
    \hline
    \multicolumn{2}{l|}{\# Users} & \multicolumn{2}{c}{35,598} & \multicolumn{2}{c}{22,363} & \multicolumn{2}{c}{19,412} & \multicolumn{2}{c}{6,040} \\
    \multicolumn{2}{l|}{\# Items} & \multicolumn{2}{c}{18,357} & \multicolumn{2}{c}{12,101} & \multicolumn{2}{c}{11,924} & \multicolumn{2}{c}{3,416} \\
    \multicolumn{2}{l|}{\# Actions} & \multicolumn{2}{c}{296,337} & \multicolumn{2}{c}{198,502} & \multicolumn{2}{c}{167,597} & \multicolumn{2}{c}{999,611} \\
    \multicolumn{2}{l|}{\# Avg. Actions/User} & \multicolumn{2}{c}{8.3} & \multicolumn{2}{c}{8.8} & \multicolumn{2}{c}{8.6} & \multicolumn{2}{c}{165.4} \\
    \multicolumn{2}{l|}{\# Avg. Actions/Item} & \multicolumn{2}{c}{16.1} & \multicolumn{2}{c}{16.4} & \multicolumn{2}{c}{14} & \multicolumn{2}{c}{292.6} \\
    \multicolumn{2}{l|}{Sparsity} & \multicolumn{2}{c}{99.95\%} & \multicolumn{2}{c}{99.93\%} & \multicolumn{2}{c}{99.93\%} & \multicolumn{2}{c}{95.16\%} \\
    \hline
    \end{tabular}}%
  \caption{Statistics of the experimented datasets.}
   \label{tab2}%
\end{table}%

\begin{table*}[t]
  \centering
  \caption{Performance comparisons of different methods. 
  The results of the best baseline are underlined in each row.
  The last column is the relative improvements compared with the best baseline results.}
      \renewcommand{\arraystretch}{1.15}
    \resizebox{\textwidth}{!}{
    \begin{tabular}{cccc|cc|cccccc|cccccccccc|cccccccc||ccr}
    \hline
    \multicolumn{1}{c}{\multirow{2}[2]{*}{DataSet}} & \multicolumn{1}{c|}{\multirow{2}[2]{*}{Metric}} & \multicolumn{1}{c|}{\multirow{2}[2]{*}{BPR}} & \multicolumn{1}{c}{\multirow{2}[2]{*}{GRU4Rec}} & \multicolumn{1}{c}{\multirow{2}[2]{*}{Caser}} & \multicolumn{1}{c|}{\multirow{2}[2]{*}{SASRec}} & \multicolumn{1}{c}{\multirow{2}[2]{*}{BERT4Rec}} & \multicolumn{1}{c}{\multirow{2}[2]{*}{S$^3$-Rec$_{\rm{MIP}}$}} & 
    \multicolumn{1}{c}{\multirow{2}[2]{*}{CL4SRec}} & \multicolumn{1}{c}{\multirow{2}[2]{*}{CoSeRec}} & \multicolumn{1}{c|}{\multirow{2}[2]{*}{DuoRec}} & \multicolumn{1}{c}{\multirow{2}[2]{*}{DSSRec}} & \multicolumn{1}{c}{\multirow{2}[2]{*}{SINE}} & \multicolumn{1}{c}{\multirow{2}[2]{*}{ICLRec}} & \multicolumn{1}{c|}{\multirow{2}[2]{*}{IOCRec}} &
    \multicolumn{1}{c}{\multirow{2}[2]{*}{ICSRec}} & \multicolumn{1}{c}{\multirow{2}[2]{*}{impro.}} \\
    
    \multicolumn{1}{c}{} & \multicolumn{1}{c|}{} & \multicolumn{1}{c|}{} & \multicolumn{1}{c}{} & \multicolumn{1}{c}{} & \multicolumn{1}{c|}{} & \multicolumn{1}{c}{} & \multicolumn{1}{c}{} & 
    \multicolumn{1}{c}{} & \multicolumn{1}{c}{} & 
    \multicolumn{1}{c|}{} & \multicolumn{1}{c}{} & \multicolumn{1}{c}{} & \multicolumn{1}{c}{}&
    \multicolumn{1}{c|}{} & \multicolumn{1}{c}{} &  \\
    \hline
    \hline
    \multicolumn{1}{c}{\multirow{6}[2]{*}{Sports}} & \multicolumn{1}{c|}{{HR@5}} & \multicolumn{1}{c|}{0.0123} & \multicolumn{1}{c}{0.0162} & \multicolumn{1}{c}{0.0154} & \multicolumn{1}{c|}{0.0214} & \multicolumn{1}{c}{0.0217} & \multicolumn{1}{c}{0.0121} & \multicolumn{1}{c}{0.0231} & \multicolumn{1}{c}{0.0290} & \multicolumn{1}{c|}{\underline{0.0312}} & \multicolumn{1}{c}{0.0209} & \multicolumn{1}{c}{0.0240} & \multicolumn{1}{c}{0.0290} & \multicolumn{1}{c|}{0.0293} & 
    \multicolumn{1}{c}{\textbf{0.0403}} & 29.17\% \\
    
    \multicolumn{1}{c}{} & \multicolumn{1}{c|}{{HR@10}} & \multicolumn{1}{c|}{0.0215} & \multicolumn{1}{c}{0.0258} & \multicolumn{1}{c}{0.0261} & \multicolumn{1}{c|}{0.0333} & \multicolumn{1}{c}{0.0359} & \multicolumn{1}{c}{0.0205} & \multicolumn{1}{c}{0.0369} & \multicolumn{1}{c}{0.0439} & \multicolumn{1}{c|}{\underline{0.0466}} & \multicolumn{1}{c}{0.0328} & \multicolumn{1}{c}{0.0389} & \multicolumn{1}{c}{0.0437} & \multicolumn{1}{c|}{0.0452}&
    \multicolumn{1}{c}{\textbf{0.0565}} & 21.24\% \\
    
    \multicolumn{1}{c}{} & \multicolumn{1}{c|}{{HR@20}} & \multicolumn{1}{c|}{0.0369} & \multicolumn{1}{c}{0.0421} & \multicolumn{1}{c}{0.0399} & \multicolumn{1}{c|}{0.0500} & \multicolumn{1}{c}{0.0604} & \multicolumn{1}{c}{0.0344} & \multicolumn{1}{c}{0.0557} & \multicolumn{1}{c}{0.0636} & \multicolumn{1}{c|}{\underline{0.0696}} & \multicolumn{1}{c}{0.0499} & \multicolumn{1}{c}{0.0610} & \multicolumn{1}{c}{0.0646} & \multicolumn{1}{c|}{0.0684}&
    \multicolumn{1}{c}{\textbf{0.0794}} & 14.10 \%\\
    
    \multicolumn{1}{c}{} & \multicolumn{1}{c|}{{NDCG@5}} & \multicolumn{1}{c|}{0.0076} & \multicolumn{1}{c}{0.0103} & \multicolumn{1}{c}{0.0114} & \multicolumn{1}{c|}{0.0144} & \multicolumn{1}{c}{0.0143} & \multicolumn{1}{c}{0.0084} & \multicolumn{1}{c}{0.0146} & \multicolumn{1}{c}{\underline{0.0196}} & \multicolumn{1}{c|}{0.0195} & \multicolumn{1}{c}{0.0139} & \multicolumn{1}{c}{0.0152} & \multicolumn{1}{c}{0.0191} & \multicolumn{1}{c|}{0.0169}&
    \multicolumn{1}{c}{\textbf{0.0283}} & 44.39 \%\\
    
    \multicolumn{1}{c}{} & \multicolumn{1}{c|}{{NDCG@10}} & \multicolumn{1}{c|}{0.0105} & \multicolumn{1}{c}{0.0142} & \multicolumn{1}{c}{0.0135} & \multicolumn{1}{c|}{0.0177} & \multicolumn{1}{c}{0.0190} & \multicolumn{1}{c}{0.0111} & \multicolumn{1}{c}{0.0191} & \multicolumn{1}{c}{\underline{0.0244}} & \multicolumn{1}{c|}{\underline{0.0244}} & \multicolumn{1}{c}{0.0178} & \multicolumn{1}{c}{0.0199} & \multicolumn{1}{c}{0.0238} & \multicolumn{1}{c|}{0.0220}&
    \multicolumn{1}{c}{\textbf{0.0335}} & 37.30\% \\
    
    \multicolumn{1}{c}{} & \multicolumn{1}{c|}{{NDCG@20}} & \multicolumn{1}{c|}{0.0144} & \multicolumn{1}{c}{0.0186} & \multicolumn{1}{c}{0.0178} & \multicolumn{1}{c|}{0.0218} & \multicolumn{1}{c}{0.0251} & \multicolumn{1}{c}{0.0146} & \multicolumn{1}{c}{0.0238} & \multicolumn{1}{c}{0.0293} & \multicolumn{1}{c|}{\underline{0.0302}} & \multicolumn{1}{c}{0.0221} & \multicolumn{1}{c}{0.0255} & \multicolumn{1}{c}{0.0291} & 
    \multicolumn{1}{c|}{0.0279}&
    \multicolumn{1}{c}{\textbf{0.0393}} & 30.13\% \\
    
   \hline
    \multicolumn{1}{c}{\multirow{6}[2]{*}{Beauty}} & \multicolumn{1}{c|}{{HR@5}} & \multicolumn{1}{c|}{0.0178} & \multicolumn{1}{c}{0.0180} & \multicolumn{1}{c}{0.0251} & \multicolumn{1}{c|}{0.0377} & \multicolumn{1}{c}{0.0360} & \multicolumn{1}{c}{0.0189} & \multicolumn{1}{c}{0.0401} & \multicolumn{1}{c}{0.0504} & \multicolumn{1}{c|}{\underline{0.0561}} & \multicolumn{1}{c}{0.0408} & \multicolumn{1}{c}{0.0354} & \multicolumn{1}{c}{0.0500} & \multicolumn{1}{c|}{0.0511}&
    \multicolumn{1}{c}{\textbf{0.0698}} & 24.42\% \\
    
    \multicolumn{1}{c}{} & \multicolumn{1}{c|}{{HR@10}} & \multicolumn{1}{c|}{0.0296} & \multicolumn{1}{c}{0.0284} & \multicolumn{1}{c}{0.0342} & \multicolumn{1}{c|}{0.0624} & \multicolumn{1}{c}{0.0601} & \multicolumn{1}{c}{0.0307} & \multicolumn{1}{c}{0.0642} & \multicolumn{1}{c}{0.0725} & \multicolumn{1}{c|}{\underline{0.0851}} & \multicolumn{1}{c}{0.0616} & \multicolumn{1}{c}{0.0612} & \multicolumn{1}{c}{0.0744} & 
    \multicolumn{1}{c|}{0.0774}&
    \multicolumn{1}{c}{\textbf{0.0960}} & 12.81\% \\
    
    \multicolumn{1}{c}{} & \multicolumn{1}{c|}{{HR@20}} & \multicolumn{1}{c|}{0.0474} & \multicolumn{1}{c}{0.0478} & \multicolumn{1}{c}{0.0643} & \multicolumn{1}{c|}{0.0894} & \multicolumn{1}{c}{0.0984} & \multicolumn{1}{c}{0.0487} & \multicolumn{1}{c}{0.0974} & \multicolumn{1}{c}{0.1034} & \multicolumn{1}{c|}{\underline{0.1228}} & \multicolumn{1}{c}{0.0894} & \multicolumn{1}{c}{0.0963} & \multicolumn{1}{c}{0.1058} & 
    \multicolumn{1}{c|}{0.1146}&
    \multicolumn{1}{c}{\textbf{0.1298}} & 5.70\% \\
    
    \multicolumn{1}{c}{} & \multicolumn{1}{c|}{{NDCG@5}} & \multicolumn{1}{c|}{0.0109} & \multicolumn{1}{c}{0.0116} & \multicolumn{1}{c}{0.0145} & \multicolumn{1}{c|}{0.0241} & \multicolumn{1}{c}{0.0216} & \multicolumn{1}{c}{0.0115} & \multicolumn{1}{c}{0.0268} & \multicolumn{1}{c}{0.0339} & \multicolumn{1}{c|}{\underline{0.0348}} & \multicolumn{1}{c}{0.0263} & \multicolumn{1}{c}{0.0213} & \multicolumn{1}{c}{0.0326} & 
    \multicolumn{1}{c|}{0.0311}&
    \multicolumn{1}{c}{\textbf{0.0494}} & 41.95\% \\
    
    \multicolumn{1}{c}{} & \multicolumn{1}{c|}{{NDCG@10}} & \multicolumn{1}{c|}{0.0147} & \multicolumn{1}{c}{0.0150} & \multicolumn{1}{c}{0.0226} & \multicolumn{1}{c|}{0.0342} & \multicolumn{1}{c}{0.0300} & \multicolumn{1}{c}{0.0153} & \multicolumn{1}{c}{0.0345} & \multicolumn{1}{c}{0.0410} & \multicolumn{1}{c|}{\underline{0.0441}} & \multicolumn{1}{c}{0.0329} & \multicolumn{1}{c}{0.0296} & \multicolumn{1}{c}{0.0403} &
    \multicolumn{1}{c|}{0.0396}& 
    \multicolumn{1}{c}{\textbf{0.0579}} & 31.29\% \\
    
    \multicolumn{1}{c}{} & \multicolumn{1}{c|}{{NDCG@20}} & \multicolumn{1}{c|}{0.0192} & \multicolumn{1}{c}{0.0186} & \multicolumn{1}{c}{0.0298} & \multicolumn{1}{c|}{0.0386} & \multicolumn{1}{c}{0.0391} & \multicolumn{1}{c}{0.0198} & \multicolumn{1}{c}{0.0428} & \multicolumn{1}{c}{0.0487} & \multicolumn{1}{c|}{\underline{0.0536}} & \multicolumn{1}{c}{0.0399} & \multicolumn{1}{c}{0.0384} & \multicolumn{1}{c}{0.0483} &
    \multicolumn{1}{c|}{0.0490}& 
    \multicolumn{1}{c}{\textbf{0.0663}} & 23.69\% \\
    
    \hline
    \multicolumn{1}{c}{\multirow{6}[2]{*}{Toys}} & \multicolumn{1}{c|}{{HR@5}} & \multicolumn{1}{c|}{0.0122} & \multicolumn{1}{c}{0.0121} & \multicolumn{1}{c}{0.0205} & \multicolumn{1}{c|}{0.0429} & \multicolumn{1}{c}{0.0371} & \multicolumn{1}{c}{0.0456} & \multicolumn{1}{c}{0.0503} & \multicolumn{1}{c}{0.0533} & \multicolumn{1}{c|}{\underline{0.0655}} & \multicolumn{1}{c}{0.0447} & \multicolumn{1}{c}{0.0385} & \multicolumn{1}{c}{0.0597} &\multicolumn{1}{c|}{0.0542} &
    \multicolumn{1}{c}{\textbf{0.0788}} & 20.30\% \\
    
    \multicolumn{1}{c}{} & \multicolumn{1}{c|}{{HR@10}} & \multicolumn{1}{c|}{0.0197} & \multicolumn{1}{c}{0.0184} & \multicolumn{1}{c}{0.0333} & \multicolumn{1}{c|}{0.0652} & \multicolumn{1}{c}{0.0524} & \multicolumn{1}{c}{0.0689} & \multicolumn{1}{c}{0.0736} & \multicolumn{1}{c}{0.0755} & \multicolumn{1}{c|}{\underline{0.0959}} & \multicolumn{1}{c}{0.0671} & \multicolumn{1}{c}{0.0631} & \multicolumn{1}{c}{0.0834} & 
    \multicolumn{1}{c|}{0.0804}&
    \multicolumn{1}{c}{\textbf{0.1055}} & 10.01\% \\
    
    \multicolumn{1}{c}{} & \multicolumn{1}{c|}{{HR@20}} & \multicolumn{1}{c|}{0.0327} & \multicolumn{1}{c}{0.0290} & \multicolumn{1}{c}{0.0542} & \multicolumn{1}{c|}{0.0957} & \multicolumn{1}{c}{0.0760} & \multicolumn{1}{c}{0.0940} & \multicolumn{1}{c}{0.0990} & \multicolumn{1}{c}{0.1037} & \multicolumn{1}{c|}{\underline{0.1293}} & \multicolumn{1}{c}{0.0942} & \multicolumn{1}{c}{0.0957} & \multicolumn{1}{c}{0.1139} &
    \multicolumn{1}{c|}{0.1132}&
    \multicolumn{1}{c}{\textbf{0.1368}} & 5.80\% \\
    
    \multicolumn{1}{c}{} & \multicolumn{1}{c|}{{NDCG@5}} & \multicolumn{1}{c|}{0.0076} & \multicolumn{1}{c}{0.0077} & \multicolumn{1}{c}{0.0125} & \multicolumn{1}{c|}{0.0245} & \multicolumn{1}{c}{0.0259} & \multicolumn{1}{c}{0.0314} & \multicolumn{1}{c}{0.0264} & \multicolumn{1}{c}{0.0370} & \multicolumn{1}{c|}{0.0392} & \multicolumn{1}{c}{0.0297} & \multicolumn{1}{c}{0.0225} & \multicolumn{1}{c}{\underline{0.0404}} &
    \multicolumn{1}{c|}{0.0297}&
    \multicolumn{1}{c}{\textbf{0.0571}} & 41.34\% \\
    
    \multicolumn{1}{c}{} & \multicolumn{1}{c|}{{NDCG@10}} & \multicolumn{1}{c|}{0.0100} & \multicolumn{1}{c}{0.0097} & \multicolumn{1}{c}{0.0168} & \multicolumn{1}{c|}{0.0320} & \multicolumn{1}{c}{0.0309} & \multicolumn{1}{c}{0.0388} & \multicolumn{1}{c}{0.0339} & \multicolumn{1}{c}{0.0442} & \multicolumn{1}{c|}{\underline{0.0490}} & \multicolumn{1}{c}{0.0369} & \multicolumn{1}{c}{0.0304} & \multicolumn{1}{c}{0.0480} &
    \multicolumn{1}{c|}{0.0381}&
    \multicolumn{1}{c}{\textbf{0.0657}} & 34.08\% \\
    
    \multicolumn{1}{c}{} & \multicolumn{1}{c|}{{NDCG@20}} & \multicolumn{1}{c|}{0.0132} & \multicolumn{1}{c}{0.0123} & \multicolumn{1}{c}{0.0221} & \multicolumn{1}{c|}{0.0397} & \multicolumn{1}{c}{0.0368} & \multicolumn{1}{c}{0.0452} & \multicolumn{1}{c}{0.0404} & \multicolumn{1}{c}{0.0513} & \multicolumn{1}{c|}{\underline{0.0574}} & \multicolumn{1}{c}{0.0437} & \multicolumn{1}{c}{0.0386} & \multicolumn{1}{c}{0.0557} &
    \multicolumn{1}{c|}{0.0464}&
    \multicolumn{1}{c}{\textbf{0.0736}} & 28.22\% \\
    
 \hline
    \multicolumn{1}{c}{\multirow{6}[2]{*}{ML-1M}} & \multicolumn{1}{c|}{{HR@5}} & \multicolumn{1}{c|}{0.0247} & \multicolumn{1}{c}{0.0806} & \multicolumn{1}{c}{0.0912} & \multicolumn{1}{c|}{0.1078} & \multicolumn{1}{c}{0.1308} & \multicolumn{1}{c}{0.1078} & \multicolumn{1}{c}{0.1142} & \multicolumn{1}{c}{0.1128} & \multicolumn{1}{c|}{\underline{0.2098}} & \multicolumn{1}{c}{0.1371} & \multicolumn{1}{c}{0.0990} & \multicolumn{1}{c}{0.1382} & 
    \multicolumn{1}{c|}{0.1796}&
    \multicolumn{1}{c}{\textbf{0.2445}} & 16.54\% \\
    
    \multicolumn{1}{c}{} & \multicolumn{1}{c|}{{HR@10}} & \multicolumn{1}{c|}{0.0412} & \multicolumn{1}{c}{0.1344} & \multicolumn{1}{c}{0.1442} & \multicolumn{1}{c|}{0.1810} & \multicolumn{1}{c}{0.2219} & \multicolumn{1}{c}{0.1952} & \multicolumn{1}{c}{0.1815} & \multicolumn{1}{c}{0.1861} & \multicolumn{1}{c|}{\underline{0.3078}} & \multicolumn{1}{c}{0.2243} & \multicolumn{1}{c}{0.1694} & \multicolumn{1}{c}{0.2273} & 
    \multicolumn{1}{c|}{0.2689}&
    \multicolumn{1}{c}{\textbf{0.3368}} & 9.42\% \\
    
    \multicolumn{1}{c}{} & \multicolumn{1}{c|}{{HR@20}} & \multicolumn{1}{c|}{0.0750} & \multicolumn{1}{c}{0.2081} & \multicolumn{1}{c}{0.2228} & \multicolumn{1}{c|}{0.2745} & \multicolumn{1}{c}{0.3354} & \multicolumn{1}{c}{0.3114} & \multicolumn{1}{c}{0.2818} & \multicolumn{1}{c}{0.2950} & \multicolumn{1}{c|}{\underline{0.4098}} & \multicolumn{1}{c}{0.3275} & \multicolumn{1}{c}{0.2705} & \multicolumn{1}{c}{0.3368} & 
    \multicolumn{1}{c|}{0.3831}&
    \multicolumn{1}{c}{\textbf{0.4518}} & 10.25\% \\
    
    \multicolumn{1}{c}{} & \multicolumn{1}{c|}{{NDCG@5}} & \multicolumn{1}{c|}{0.0159} & \multicolumn{1}{c}{0.0475} & \multicolumn{1}{c}{0.0565} & \multicolumn{1}{c|}{0.0681} & \multicolumn{1}{c}{0.0804} & \multicolumn{1}{c}{0.0616} & \multicolumn{1}{c}{0.0705} & \multicolumn{1}{c}{0.0692} & \multicolumn{1}{c|}{\underline{0.1433}} & \multicolumn{1}{c}{0.0898} & \multicolumn{1}{c}{0.0586} & \multicolumn{1}{c}{0.0889} & 
    \multicolumn{1}{c|}{0.1201}&
    \multicolumn{1}{c}{\textbf{0.1710}} & 19.33\% \\
    
    \multicolumn{1}{c}{} & \multicolumn{1}{c|}{{NDCG@10}} & \multicolumn{1}{c|}{0.0212} & \multicolumn{1}{c}{0.0649} & \multicolumn{1}{c}{0.0734} & \multicolumn{1}{c|}{0.0948} & \multicolumn{1}{c}{0.1097} & \multicolumn{1}{c}{0.0917} & \multicolumn{1}{c}{0.0920} & \multicolumn{1}{c}{0.0915} & \multicolumn{1}{c|}{\underline{0.1749}} & \multicolumn{1}{c}{0.1179} & \multicolumn{1}{c}{0.0812} & \multicolumn{1}{c}{0.1175} & 
    \multicolumn{1}{c|}{0.1487}&
    \multicolumn{1}{c}{\textbf{0.2007}} & 14.75\% \\
    
    \multicolumn{1}{c}{} & \multicolumn{1}{c|}{{NDCG@20}} & \multicolumn{1}{c|}{0.0297} & \multicolumn{1}{c}{0.0834} & \multicolumn{1}{c}{0.0931} & \multicolumn{1}{c|}{0.1156} & \multicolumn{1}{c}{0.1384} & \multicolumn{1}{c}{0.1204} & \multicolumn{1}{c}{0.1170} & \multicolumn{1}{c}{0.1247} & \multicolumn{1}{c|}{\underline{0.2007}} & \multicolumn{1}{c}{0.1440} & \multicolumn{1}{c}{0.1066} & \multicolumn{1}{c}{0.1450} &
    \multicolumn{1}{c|}{0.1775}&
    \multicolumn{1}{c}{\textbf{0.2297}} & 14.45\% \\
 \hline
    \end{tabular}}%
      \label{tab:main}
\end{table*}%
\section{Experiments}
In this section, we present our experimental setup and empirical results. 
Our experiments are designed to investigate the following research questions(\textbf{RQ}s):
\begin{itemize}[leftmargin=*]
    \item \textbf{RQ1:} How does ICSRec perform compared to the State-Of-The-Art (SOTA) SR models?
    \item \textbf{RQ2:} How effective are the key model components (e.g., intent representation learning, FNM) in ICSRec?
    \item \textbf{RQ3:} How does the robustness (e.g., hyper-parameters sensitivity, against noise interactions) of ICSRec?
\end{itemize}

\subsection{Experimental Settings}
\noindent\textbf{Datasets.} We evaluate the effectiveness of our proposed model on four datasets from two real-world applications:
\begin{itemize}[leftmargin=*]
    \item \textbf{Amazon\footnote{\url{http://jmcauley.ucsd.edu/data/amazon/}}:} 
    A series of datasets that collect user reviews on products from $Amazon.com$. 
    The dataset can be divided into many subsets according to the various product categories, which have relatively short sequence lengths. 
    In this paper, we pick \textbf{Sports}, \textbf{Beauty}, and \textbf{Toys} as three different experimental datasets from the Amazon dataset.
    \item \textbf{MovieLens\footnote{\url{https://grouplens.org/datasets/movielens/1m/}}:} 
    It contains users' behavior logs on movies, which have very long sequences. 
    We pick the version MovieLens-1M (\textbf{ML-1M}), which includes 1 million user ratings, as an experimental dataset.
\end{itemize}
To ensure the data quality, users or items appearing less than five times are removed~\cite{SASRec,BERT4Rec}. 
For all datasets, we aggregate each user's interaction records and sort them by the action timestamps in chronological order. 
For evaluation purposes, we follow ~\cite{ICLRec,DuoRec} to split the data into training, validation, and testing datasets based on timestamps given in the datasets. 
Specifically, the last item is used as the label for testing, the second-to-last item is used as the label for validating, and the others for training. 
Table~\ref{tab2} presents the detailed statistics of the four datasets.

\begin{table}[t]
  \centering
      \caption{The HR@20 and NDCG@20 performances achieved by ICSRec variants and SASRec on four datasets.}
    \renewcommand{\arraystretch}{1.5}
  \resizebox{1.0\linewidth}{!}{
    \begin{tabular}{l|cc|cc|cc|cc}
    \hline
    \multicolumn{1}{c|}{\multirow{3}[4]{*}{Model}} & \multicolumn{8}{c}{Dataset} \\
\cline{2-9}          & \multicolumn{2}{c|}{Sports} & \multicolumn{2}{c|}{Beauty} & \multicolumn{2}{c|}{Toys} & \multicolumn{2}{c}{ML-1M} \\
          & \multicolumn{1}{c}{HR} & \multicolumn{1}{c|}{NDCG} & \multicolumn{1}{c}{HR} & \multicolumn{1}{c|}{NDCG} & \multicolumn{1}{c}{HR} & \multicolumn{1}{c|}{NDCG} & \multicolumn{1}{c}{HR} & \multicolumn{1}{c}{NDCG} \\
    \hline
     (A) ICSRec & \textbf{0.0794} & \textbf{0.0393} & \textbf{0.1298} & \textbf{0.0663} & \textbf{0.1368} & \textbf{0.0736} & \textbf{0.4518} & \textbf{0.2297} \\
     (B) w/o CICL  & 0.0754 & 0.0374 & 0.1268 & 0.0627 & 0.1293 & 0.0701 & 0.4468 & 0.2174 \\
     (C) w/o FICL & 0.0746 & 0.0368 & 0.1252 & 0.0616 & 0.1346 & 0.0716 & 0.4455 & 0.2162 \\
     (D) w/o FNM & 0.0786 & 0.0384 & 0.1290 & 0.0658 & 0.1336 & 0.0718 & 0.4315 & 0.2157 \\
     \hline
    (E) SASRec & 0.0500 & 0.0218 & 0.0894 & 0.0386 & 0.0957 & 0.0397 & 0.2745 & 0.1156 \\
     \hline
    \end{tabular}}%
  \label{tab:ablation}%
\end{table}%

\noindent\textbf{Baseline Models.} We compare ICSRec with the following representative SR models:
\begin{itemize}[leftmargin=*]
    \item 
    \textbf{Non-sequential model:}
    \textbf{BPR}~\cite{BPR} first employs Bayesian Personalized Ranking (BPR) loss to optimize the matrix factorization model. 
    \item
    \textbf{General sequential models:} 
    \textbf{GRU4Rec}~\cite{GRU4Rec} introduces a Gated Recurrent Unit (GRU) to model sequences and first leverages Recurrent Neural Networks (RNN) for SR. 
    \textbf{Caser}~\cite{Caser} first introduces Convolutional Neural Network (CNN) to SR, which leverages both horizontal and vertical convolution to model the sequence. 
    \textbf{SASRec}~\cite{SASRec} first utilizes the attention mechanism to model sequences, which greatly improves the performance of SR. 
    \item
    \textbf{SSL-based sequential models:} 
    \textbf{BERT4Rec}~\cite{BERT4Rec}  first introduces BERT~\cite{BERT} to model the sequence, which leverages MIP (Mask Item Predict) task to capture the potential relationships between items and sequences. 
    \textbf{S$^{3}$-Rec$_{\rm{MIP}}$}~\cite{S3Rec} first introduces self$\mbox{-}$supervised learning to capture the potential relationships between items. 
    And since we have no attribute information, only the MIP task, called S$^{3}$-Rec$_{\rm{MIP}}$, is used for training. 
    \textbf{CL4SRec}~\cite{CL4SRec} first integrates data augmentation and contrastive learning to SR, which further improves the robustness of the model to noise and sparsity 
    \textbf{CoSeRec}~\cite{CoSeRec} further improves CL4SRec by introducing two more robust data augmentation. 
    \textbf{DuoRec}~\cite{DuoRec} introduces a model augmentation method and a supervised sampling strategy for the first time. 
    \item
    \textbf{Intent-guided sequential models:} 
    \textbf{DSSRec}~\cite{DSSRec} first introduces a novel seq2seq training strategy and an intention-disentan-glement layer for SR. 
    \textbf{SINE}~\cite{SINE} designs an adaptive interest aggregation module to model users' multiple interests for SR. 
    \textbf{ICLRec}~\cite{ICLRec} learns users' latent intents from the behavior sequences via clustering and integrates the learned intents into the model via an auxiliary contrastive loss. 
    \textbf{IOCRec}~\cite{IOCRec} first introduces intent-level contrastive learning for denoising problems of the SR task.
\end{itemize}

\begin{figure}[t]
        \begin{minipage}[t]{0.49\linewidth}
        \centerline{\includegraphics[width=1.0\textwidth]{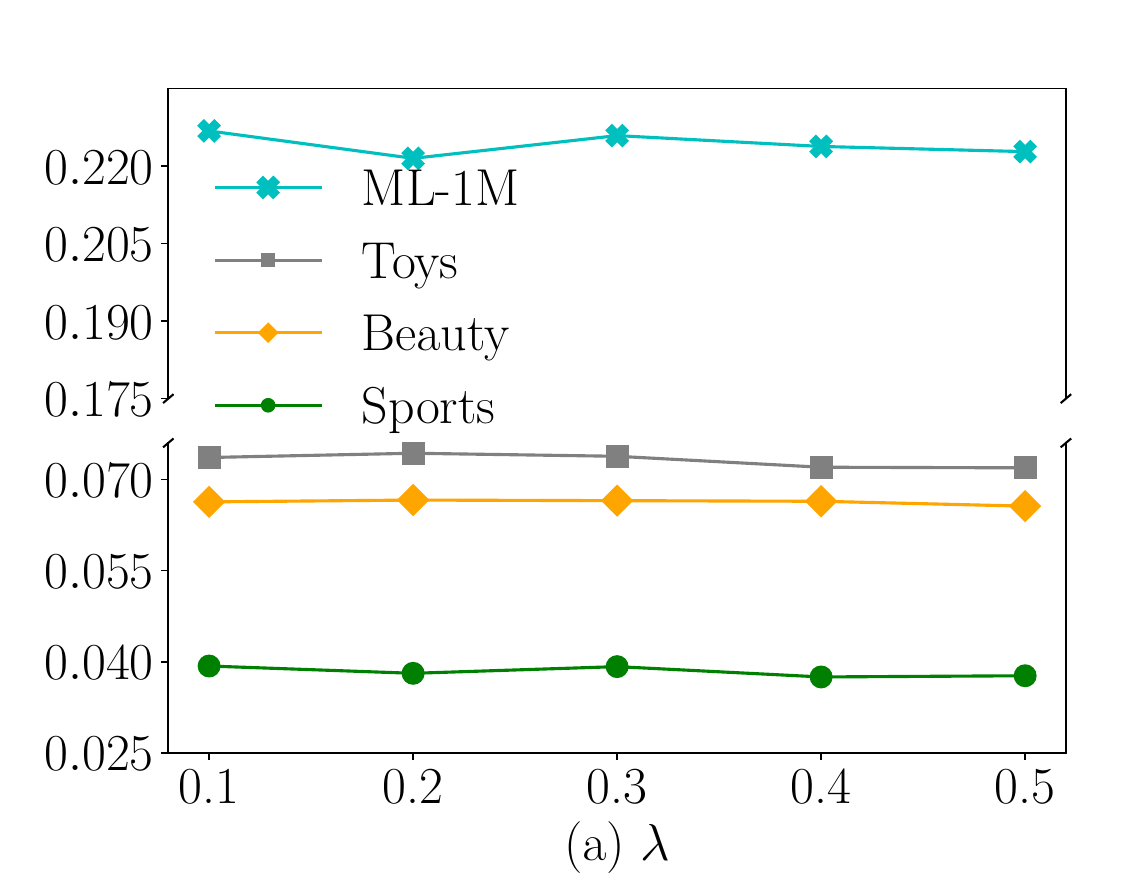}
        }
     \end{minipage}
        \begin{minipage}[t]{0.49\linewidth}
        \centerline{\includegraphics[width=1.0\textwidth]{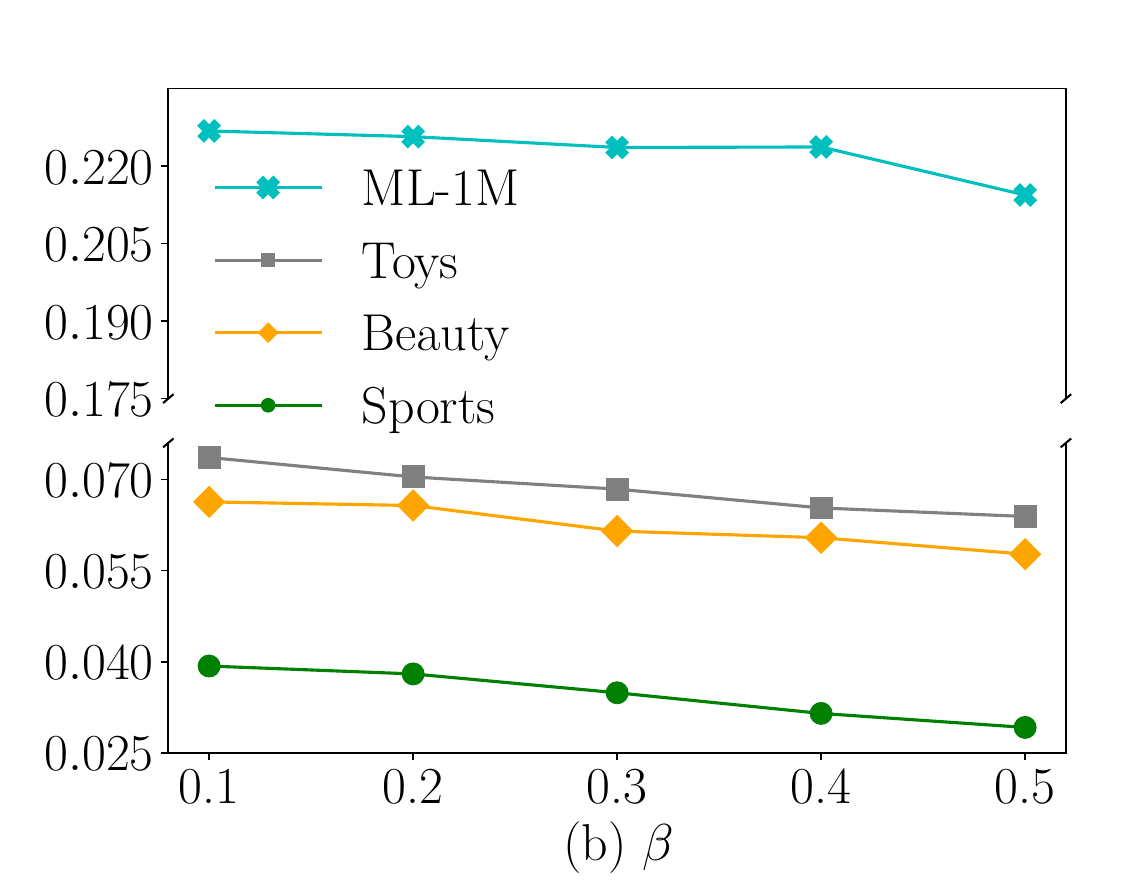}}
     \end{minipage}
            \begin{minipage}[t]{0.49\linewidth}
    \centerline{\includegraphics[width=1.0\textwidth]{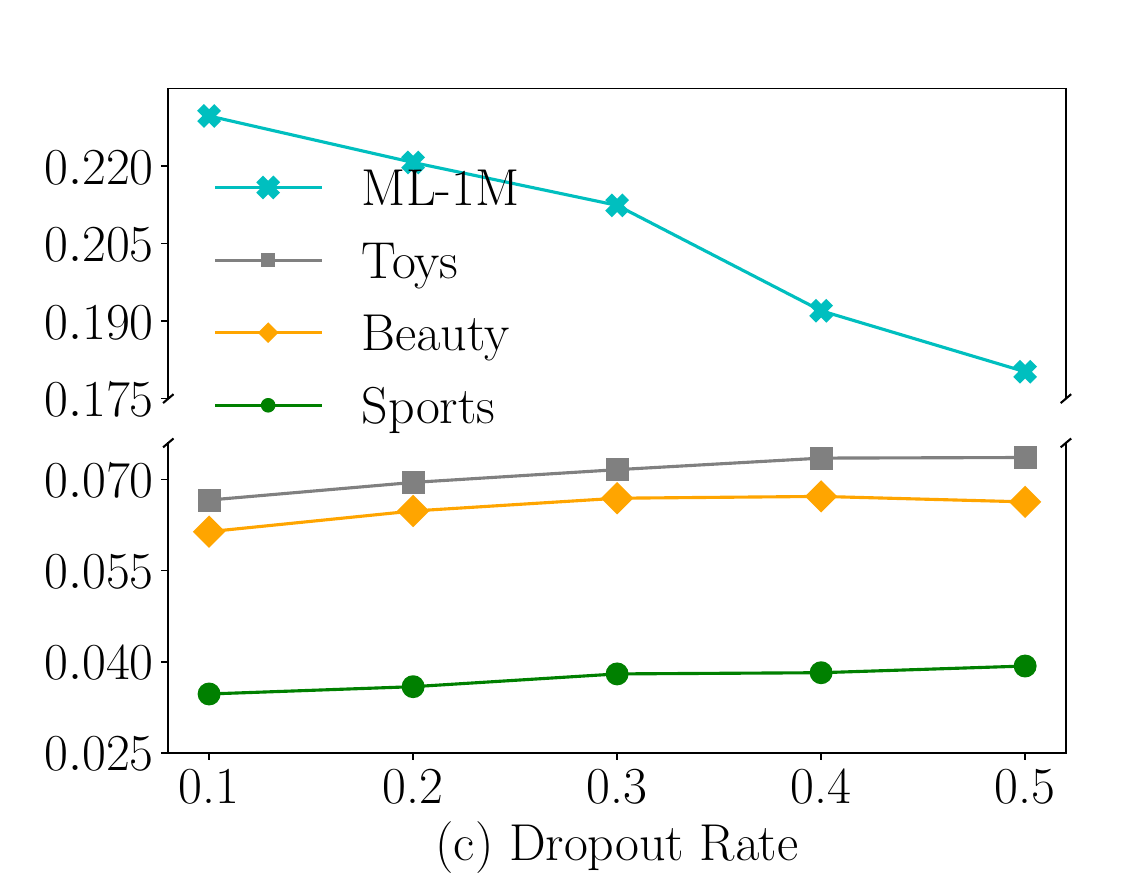}}
     \end{minipage}
    \begin{minipage}[t]{0.49\linewidth}
    \centerline{\includegraphics[width=1.0\textwidth]{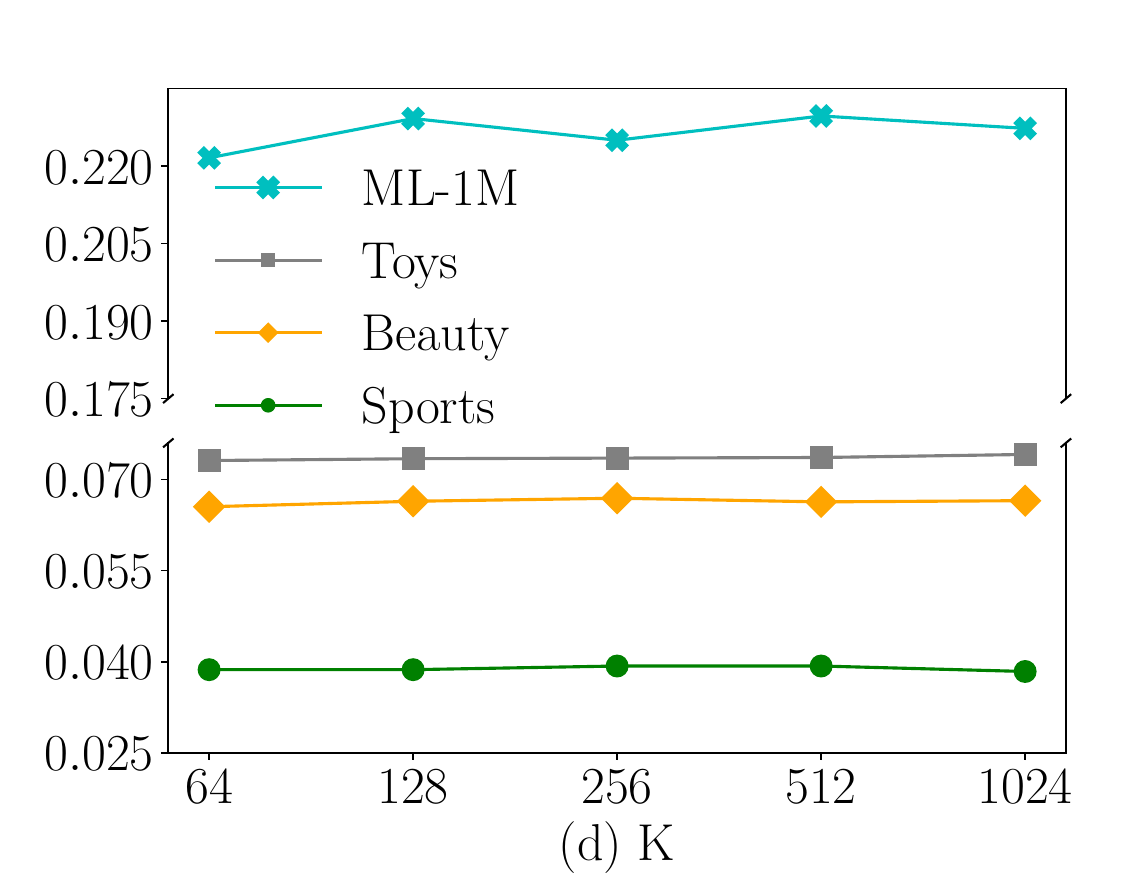}}
     \end{minipage}

     
\caption{Performances of ICSRec w.r.t. different hyper-parameters ($\lambda$, $\beta$, dropout rate and $K$). on NDCG@20.}
\label{fig_b}
\end{figure}
\noindent\textbf{Evaluation Metrics.} 
We follow~\cite{Metric1,Metric2} to rank the whole item set without negative sampling and use two widely$\mbox{-}$used evaluation metrics to evaluate the model, including Hit Ratio @$k$ (HR@$k$) and Normalized Discounted Cumulative Gain @$k$ (NDCG@$k$) where $k \in \{5,10,20\}$. 

\noindent\textbf{Implementation Details.} 
The implementations of Caser, S$^{3}$-Rec, BERT4Rec, CoSeRec, ICLRec, DuoRec, and IOCRec are provided by the authors. 
BPR, GRU4Rec, SASRec, CL4SRec, DSSRec, and SINE are implemented based on public resources. 
All parameters in these methods are used as reported in their papers and the optimal settings are chosen based on the model performance on validation data. 
For ICSRec, the number of the self-attention blocks and attention heads is set as 2. 
The batch size is set to 256. 
We set $d$ as 64, $n$ as 50, $\tau$ as 1.0. 
$\lambda$ and $\beta$ are selected from $\{0.1, 0.2, 0.3, 0.4, 0.5\}$. 
We use Adam~\cite{adam} as the optimizer and set the learning rate to $10^{-3}$. 
The number of clusters is set in the range of $\{64, 128, 256, 512, 1024\}$ and the dropout rate in a range of $\{0.1, 0.2, 0.3, 0.4, 0.5\}$. 
We train the model with an early stopping strategy based on the performance of validation data. 
All experiments are implied on NVIDIA GeForce RTX 2080 Ti GPU. 

\begin{figure}[t]
        \begin{minipage}[t]{0.49\linewidth}
        \centerline{\includegraphics[width=1.0\textwidth]{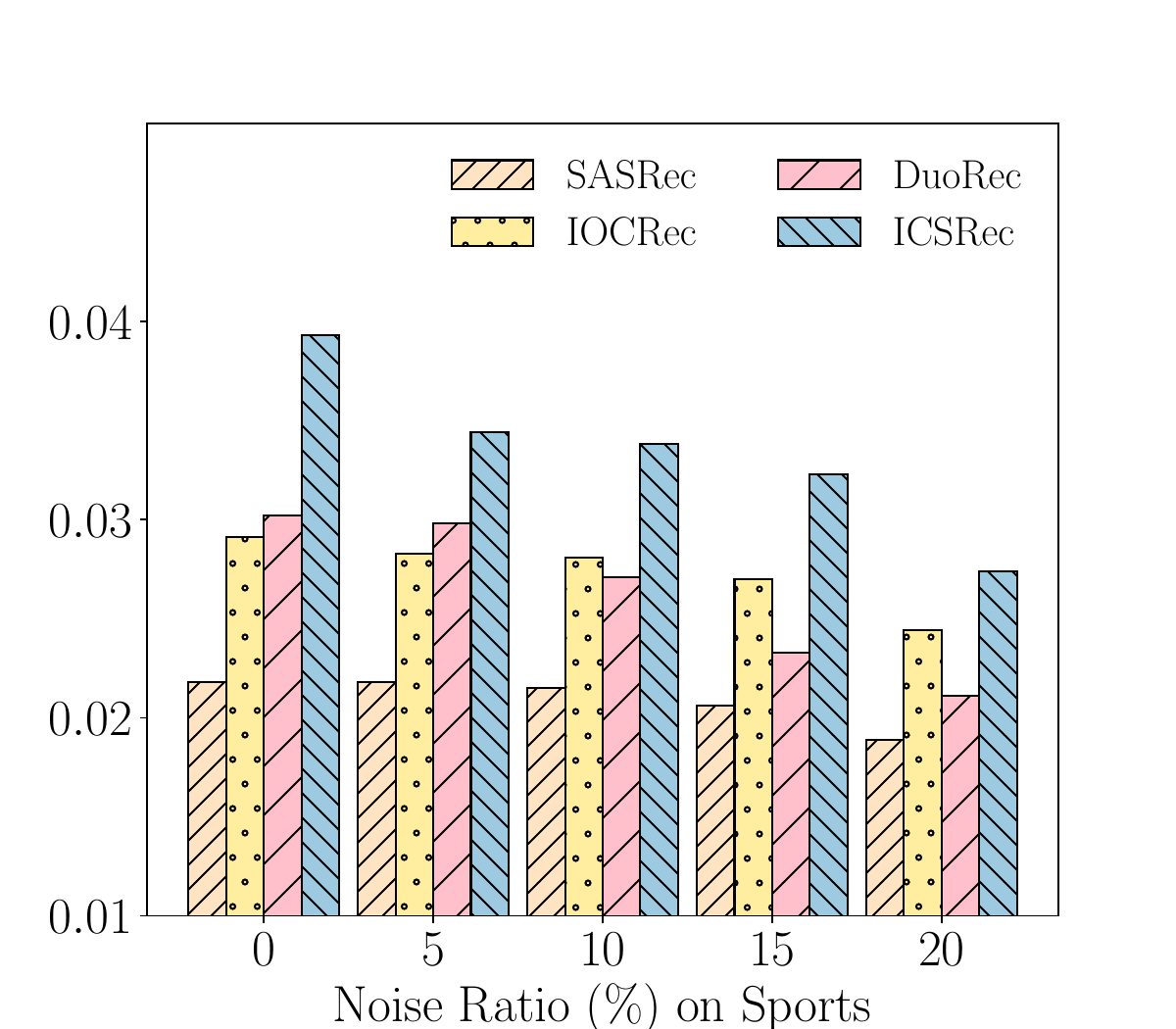}}
     \end{minipage}
        \begin{minipage}[t]{0.49\linewidth}
        \centerline{\includegraphics[width=1.0\textwidth]{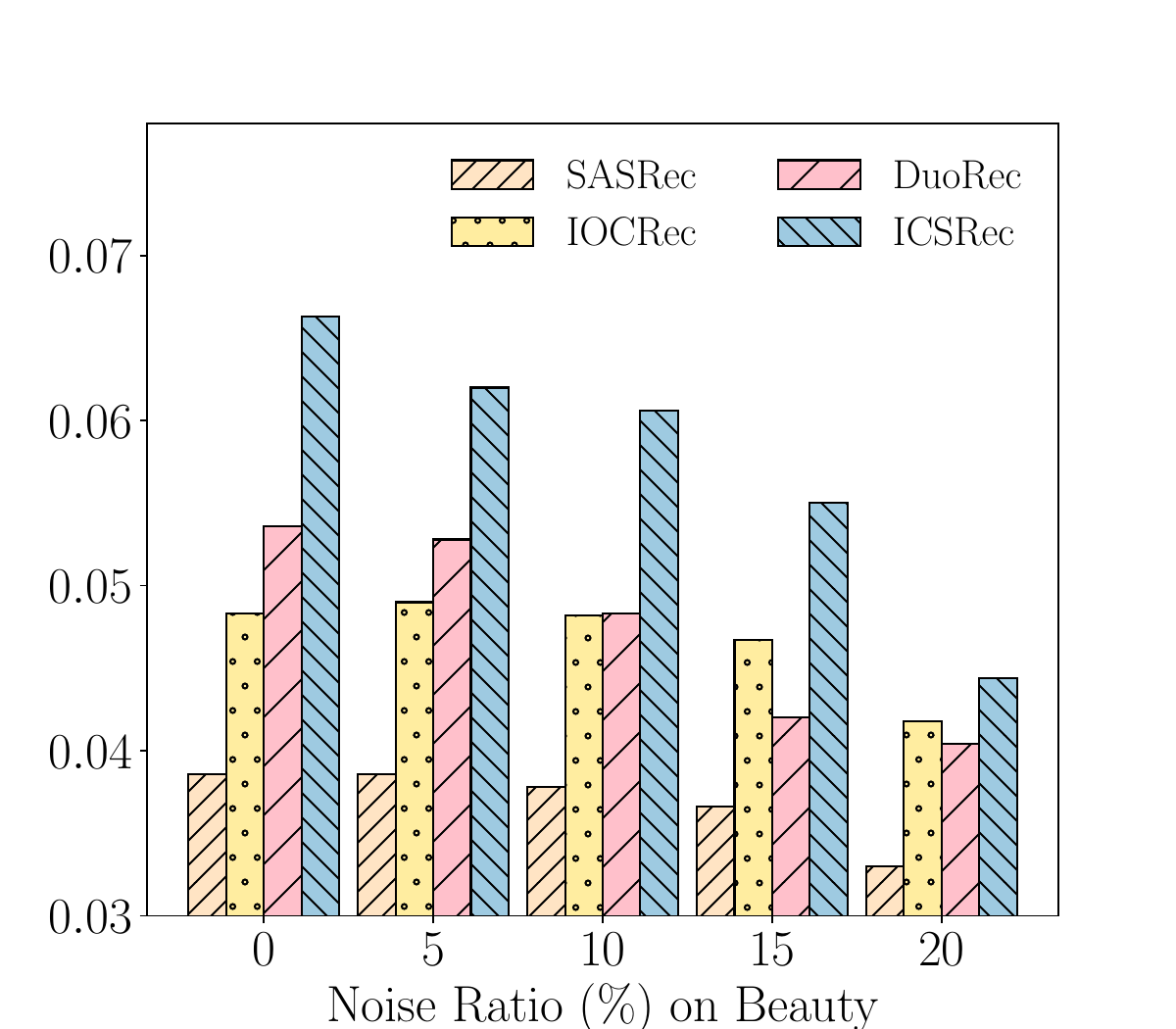}}
     \end{minipage}
    \begin{minipage}[t]{0.49\linewidth}
        \centerline{\includegraphics[width=1.0\textwidth]{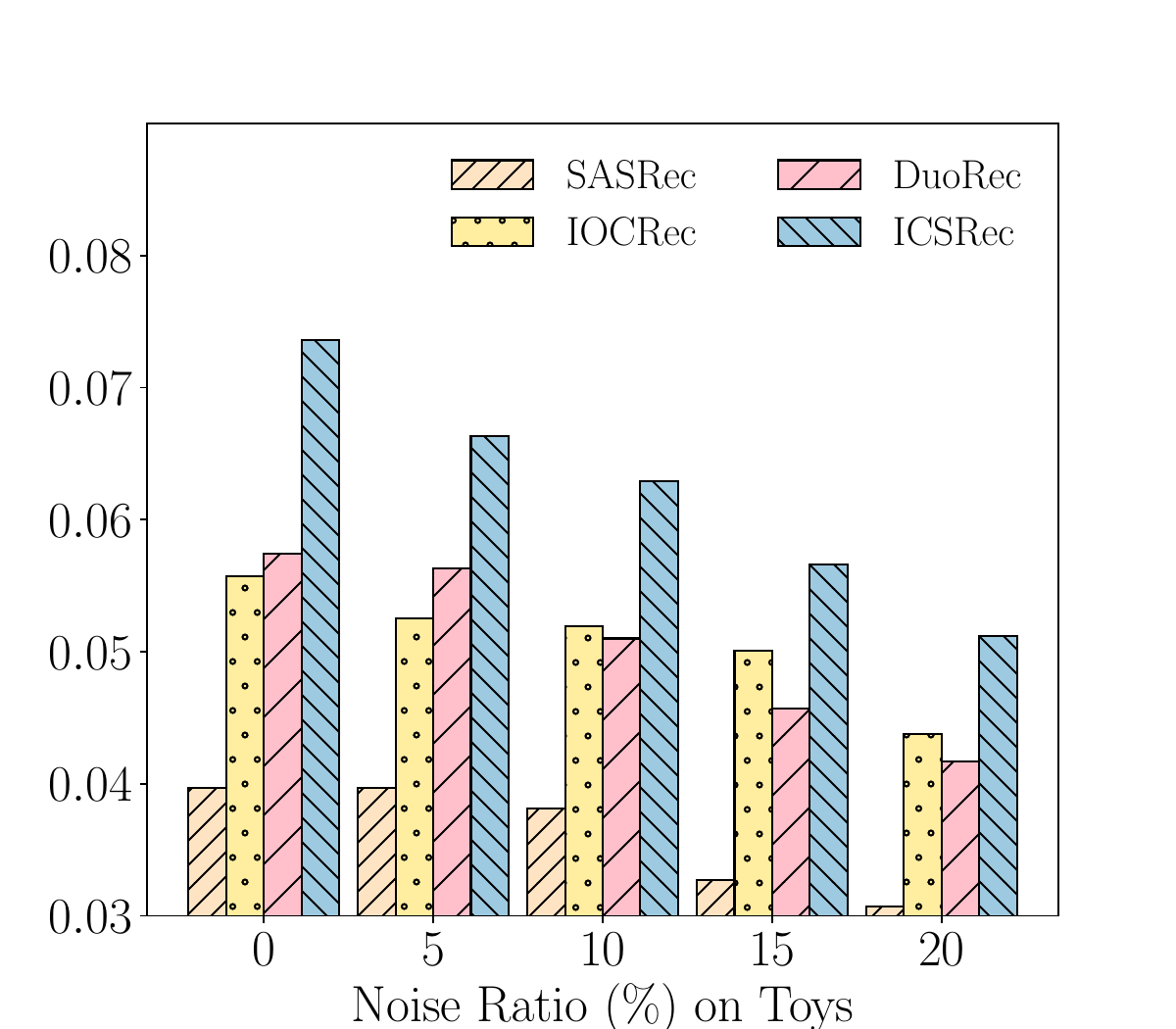}}
     \end{minipage}
    \begin{minipage}[t]{0.49\linewidth}
        \centerline{\includegraphics[width=1.0\textwidth]{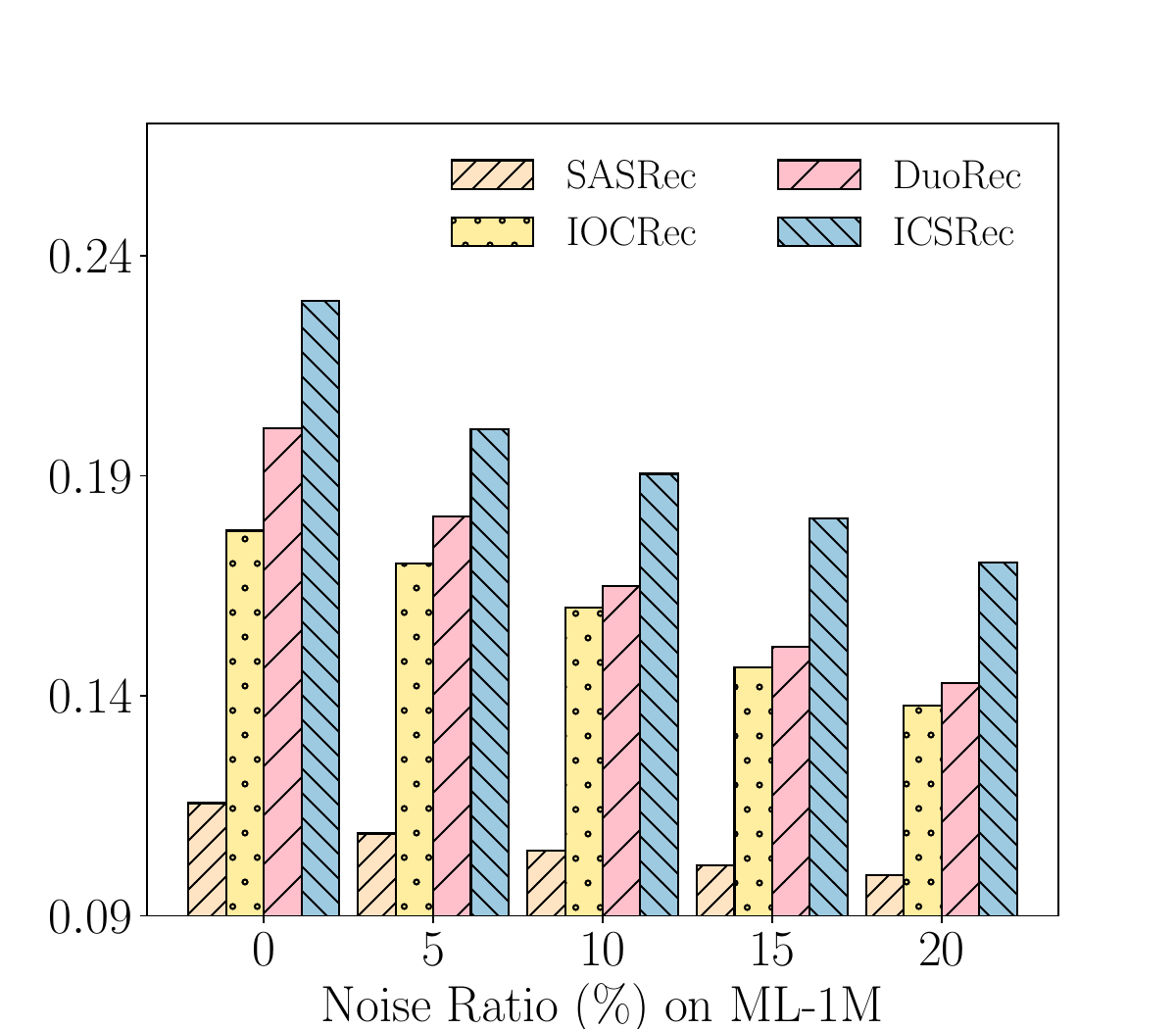}}
     \end{minipage}
\caption{Performance comparison w.r.t. noise ratio on
four datasets. The bar chart shows the performance in NDCG@20.}
\label{noise_exp}
\end{figure}

\subsection{ Performance Comparison (RQ1)}
We compare the performance of all baselines with ICSRec. 
Table~\ref{tab:main} shows the experimental results of all the compared models on four datasets, and the following findings can be seen through it:
\begin{itemize}[leftmargin=*]
    \item 
    Models based on the attention mechanism are better than models based on other networks, such as GRU4Rec and Caser, which demonstrates the superiority of the attention mechanism in modeling sequential tasks. 
    In addition, the self-supervised based models perform more effectively than classical models, such as SASRec. 
    Among them, different from BERT4Rec and S$^{3}$-Rec$_{\rm{MIP}}$ that use MIP tasks to train the model, CL4SRec, CoSeRec, and DuoRec utilize data augmentation and Contrastive Learning (CL) for training, which leads to generally better results than BERT4Rec and S$^{3}$-Rec$_{\rm{MIP}}$. 
    That indicates the CL paradigm may generate more expressive embeddings for users and items by maximizing the mutual information. 
    \item 
    The intent-based models perform better than most self-supervised models, such as BERT4Rec, S$^{3}$-Rec$_{\rm{MIP}}$, and CL4SRec. 
    Among them, ICLRec and IOCRec achieve better results in this group. 
    Different from DSSRec and SINE, ICLRec and IOCRec utilize a contrastive SSL task to learn the intent representation for SR, and thus improve the performance by a large margin. 
    But they are not as effective as DuoRec, probably because introducing stochastic data augmentation destroys the intention of the original sequences and thus reduces the performance~\cite{MISS}. 
    That motivates us to further investigate how to better combine CL and intent modeling for SR. 
    \item 
    Benefiting from introducing intent supervisory signal for intent representation learning, ICSRec significantly outperforms other methods on all metrics across the different datasets. 
    For instance, ICSRec improves over the best baseline result w.r.t. NDCG by 23.69-44.39\% and 14.45-19.33\% on three sparse datasets and the dense dataset ML-1M, respectively. 
    The reason may be that the user's interaction sequences are longer in dense dataset ML-1M, which makes the users' intentions more diverse and less easy to distinguish. 
\end{itemize}

\begin{table}[t]
    \centering
        \caption{The HR@20 and NDCG@20 performances of GRU4Rec, SASRec, and ICSRec with different sequence encoders.}
    \renewcommand{\arraystretch}{1.5}
    \resizebox{1.0\linewidth}{!}{
    \begin{tabular}{l|cc|cc|cc|cc}
    \hline
    \multicolumn{1}{c|}{\multirow{3}[4]{*}{Model}} & \multicolumn{8}{c}{Dataset} \\
\cline{2-9}          & \multicolumn{2}{c|}{Sports} & \multicolumn{2}{c|}{Beauty} & \multicolumn{2}{c|}{Toys} & \multicolumn{2}{c}{ML-1M} \\
          & \multicolumn{1}{c}{HR} & \multicolumn{1}{c|}{NDCG} & \multicolumn{1}{c}{HR} & \multicolumn{1}{c|}{NDCG} & \multicolumn{1}{c}{HR} & \multicolumn{1}{c|}{NDCG} & \multicolumn{1}{c}{HR} & \multicolumn{1}{c}{NDCG} \\
    \hline
     (A) GRU4Rec & 0.0421 & 0.0186 & 0.0478 & 0.0186 & 0.0290 & 0.0123 & 0.2081 & 0.0834 \\
     (B) ICSRec$_{GRU}$  & 0.0595 & 0.0280 & 0.1014 & 0.0506 & 0.0950 & 0.0510 & 0.4045 & 0.1964 \\
         \hline
    (C) SASRec & 0.0500 & 0.0218 & 0.0894 & 0.0386 & 0.0957 & 0.0397 & 0.2745 & 0.1156 \\
     (D) ICSRec$_{SAS}$ & \textbf{0.0794} & \textbf{0.0393} & \textbf{0.1298} & \textbf{0.0663} & \textbf{0.1368} & \textbf{0.0736} & \textbf{0.4518} & \textbf{0.2297} \\
     \hline
    \end{tabular}}%
    \label{tab:encoder}%
    \end{table}%

\begin{table*}[t]
  \centering
    \renewcommand{\arraystretch}{1.5}
  \resizebox{1.0\linewidth}{!}{
    \begin{tabular}{l|c|c|c|c|c|c|c|c|c|c|c}
    \hline
    User1 & \{185, T\} & \{3307, D$\&$Co\} & \{1287, D$\&$A\} & \{2283, D\} & \{1363, D\} & \{1252, T\} & \{1276, D$\&$Co\} & \{3007, Do\} & \{627, D$\&$T\} & \{3791, D\} &  \multicolumn{1}{c}{\multirow{2}[4]{*}{ \{ 2579, D\} }} \\
\cline{1-11}    User2 & \{3889, A\} & \{1991, H\} & \{3525, Co\} & \{2335, Co\} & \{2920, D\} & \{265, D\} & \{3773, Co\} & \{381, D\} & \{3246, D\} & \{3028, Co\} &  \\
    \hline
    User3 & \{546, A$\&$C\} & \{1431, A$\&$Co\} & \{2412, A$\&$D\} & \{2153, A\} & \{694, A\} & \{3113, A$\&$T\} & \{1497, A\} & \{1772, A$\&$Co\} & \{2487, A$\&$Co\} & \{2568, A\} &  \multicolumn{1}{c}{\multirow{2}[4]{*}{\{2720, A$\&$C\}}} \\
\cline{1-11}    User4 & \{2142, C$\&$Co\} & \{2162, C\} & \{126, C\} & \{546, C$\&$A\} & \{631, C$\&$M\} & \{87, C$\&$Co\} & \{1702, C$\&$Co\} & \{575, C$\&$Co\} & \{3054, C\} & \{1822, C$\&$Co\} &  \\
    \hline
    \end{tabular}}%
    \caption{The case study explains the motivation of our proposed model. The digit means the movie id from the ML-1M datasets. The capital letter means the film genre (A: Action; Co: Comedy; C: Children; D: Drama; T: Thriller; M: Musical; Do: Documentary; H: Horror). User1 and User2, User3 and User4 have the same target movie.}
    \label{tab_intent}%
\end{table*}%
\subsection{Ablation Study (RQ2)}
To analyze the effectiveness of each component of our model, we conduct several ablation experiments about ICSRec, where w/o denotes without. 
(A) denotes ICSRec, (B) removes the coarse-grain intent learning module by setting $\lambda$ to 0 in Eq.(\ref{eq11}), (C) removes the fine-grain intent learning module by setting $\beta$ to 0 in Eq.(\ref{eq11}), and (D) removes the false negative mask component in Eq.(\ref{eq6}). (E) denotes the base encoder SASRec~\cite{SASRec}. 

Tabel~\ref{tab:ablation} summarizes HR@20 and NDCG@20 performances of ICSRec variants and SASRec on four datasets. 
From the table, we can find that ICSRec achieves the best results on all datasets, which indicates all components are effective for our framework. 
By comparing (A) with (B) and (C), we find that CICL and FICL could significantly improve the model accuracy, which is consistent with our statements. 
By comparing (B) and (C), it can be observed that FICL is more efficient than CICL. 
By comparing (A) and (D), we can find that the mask of the false negative sample could improve the performance. 
By comparing (A) and (E), we can find that ICSRec outperforms the backbone model SASRec on four datasets. 
This indicates that leveraging the intent supervisory signals, obtained from cross subsequences, for intent representation learning can help improve SR performance.

\begin{figure}[t]
        \begin{minipage}[t]{0.49\linewidth}
        \centerline{\includegraphics[width=1.0\textwidth]{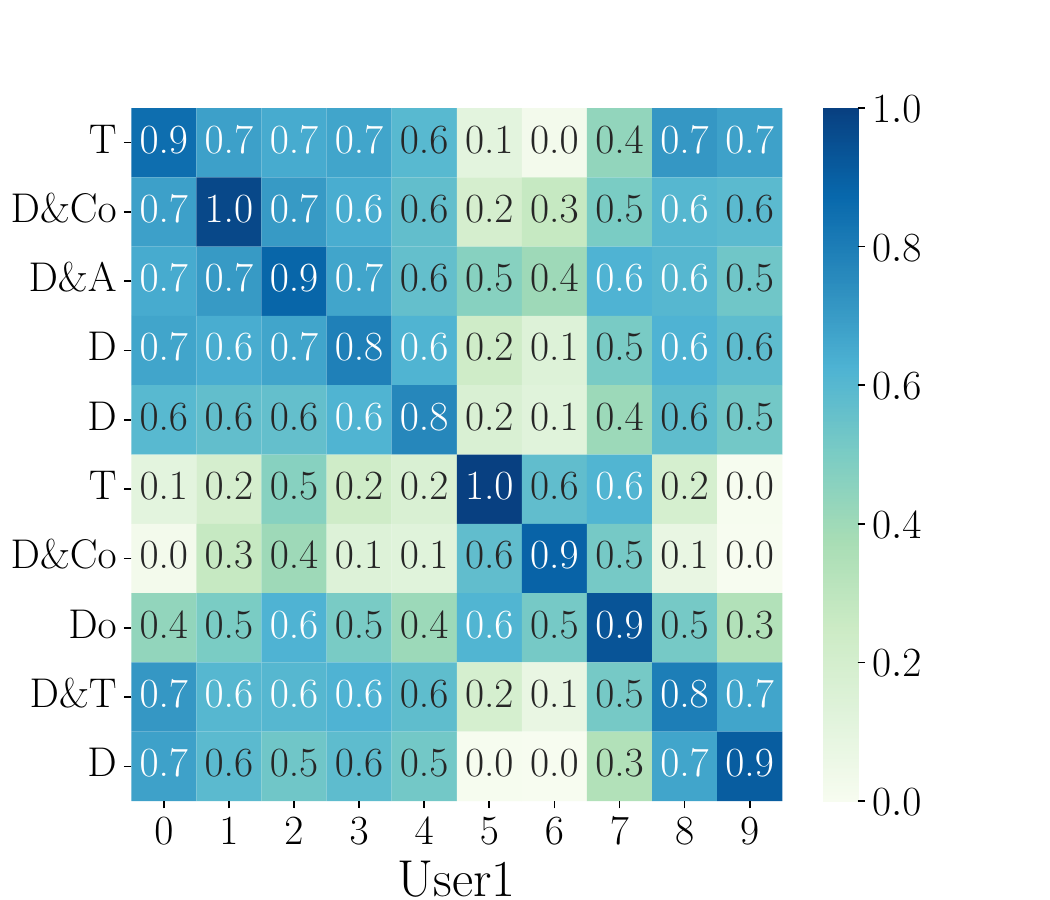}}
        \end{minipage}
        \begin{minipage}[t]{0.49\linewidth}
        \centerline{\includegraphics[width=1.0\textwidth]{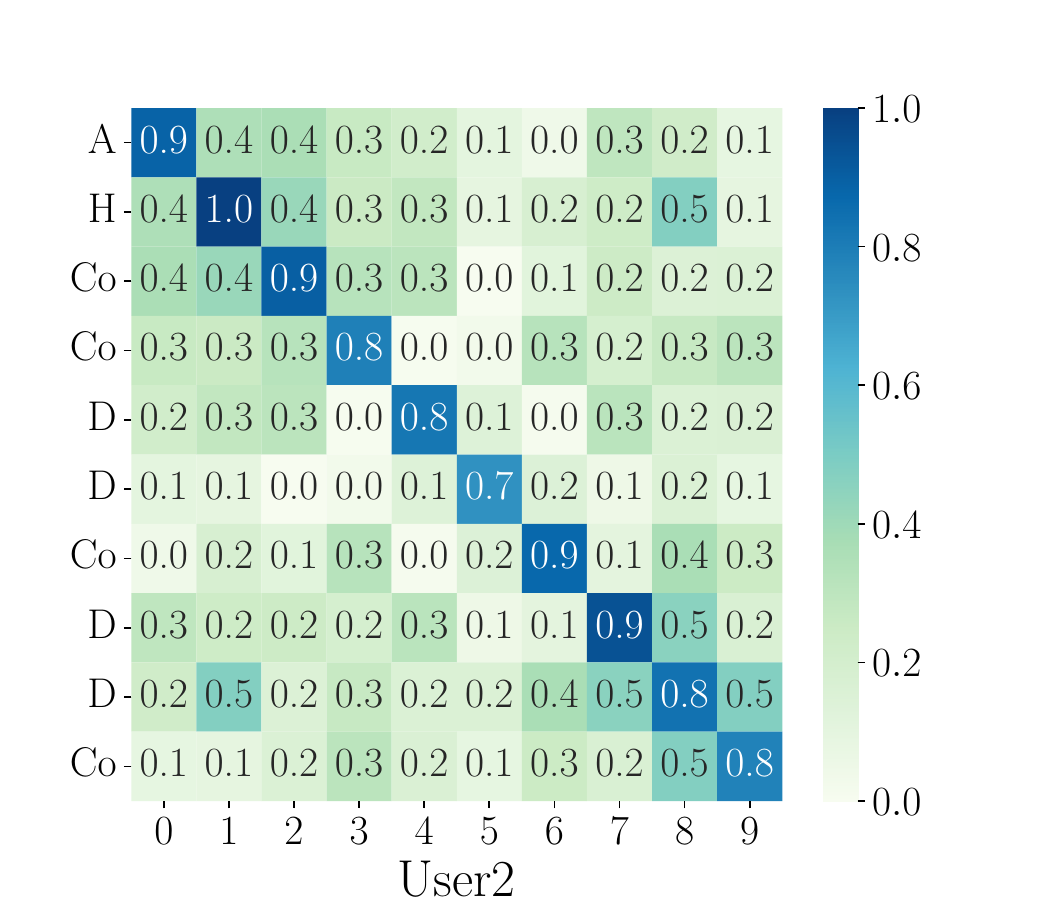}}
     \end{minipage}
        \begin{minipage}[t]{0.49\linewidth}
        \centerline{\includegraphics[width=1.0\textwidth]{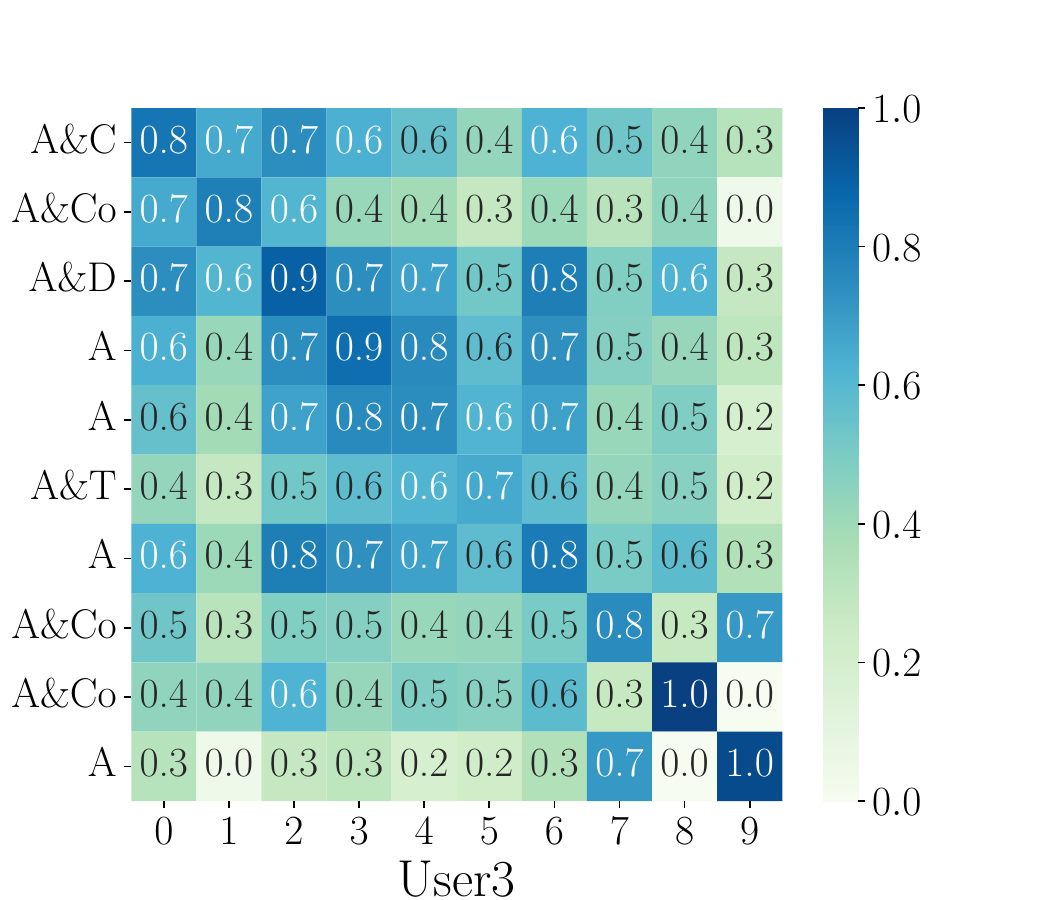}}
        \end{minipage}
        \begin{minipage}[t]{0.49\linewidth}
        \centerline{\includegraphics[width=1.0\textwidth]{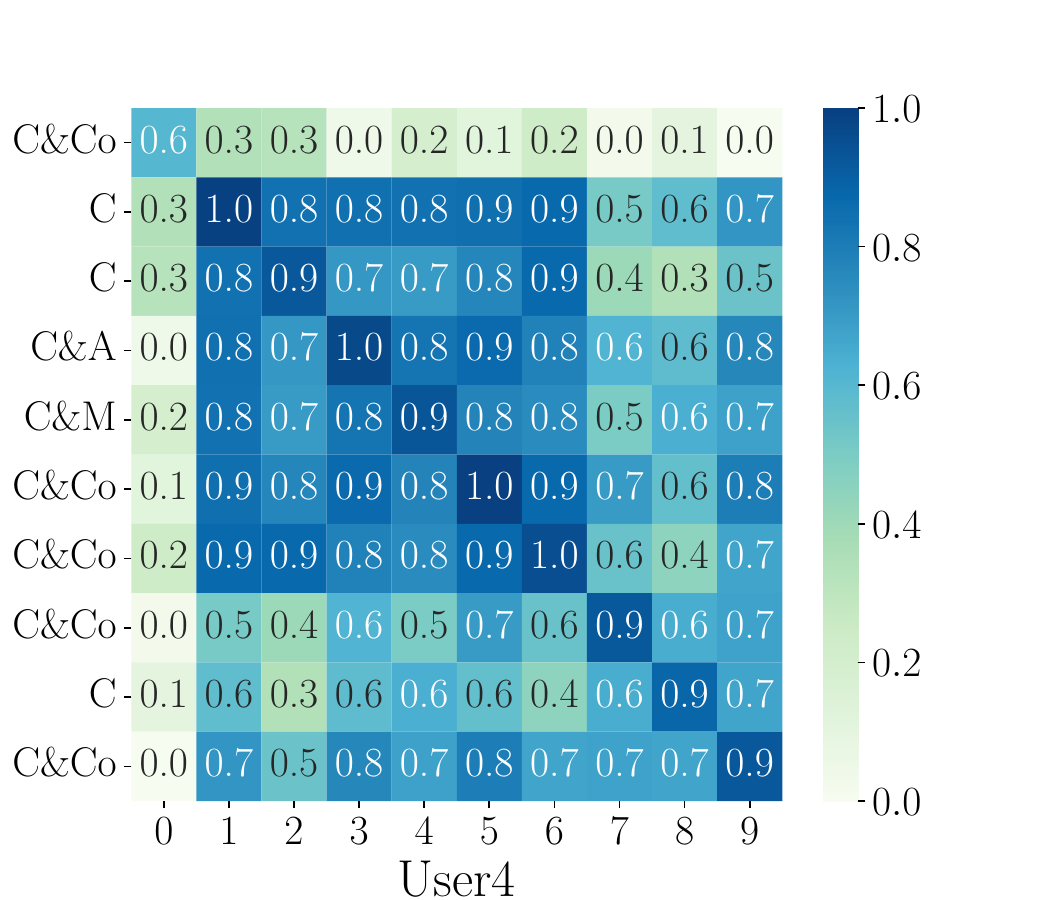}}
     \end{minipage}
       \caption{Visualization of the intents heat map for four users.}
         \label{intent}
\end{figure}
\subsection{Further Analysis (RQ3)}
We conduct various experiments on four datasets to verify the robustness of ICSRec. For all models in the following experiments, we only change one variable at one time while keeping other hyper-parameters optimal.

\noindent\textbf{Impact of $\mathcal{L}_{cicl}$ and $\mathcal{L}_{ficl}$ weights.}
The final loss function of ICSRec in Eq.(\ref{eq11}) is a multi$\mbox{-}$task learning loss. 
Figure~\ref{fig_b}(a) and (b) shows the effect of assigning different weights to $\lambda$ and $\beta$ on the model, respectively. 
We observe that the performance of ICSRec gets peak value to different $\beta$ and $\lambda$, which demonstrates the effectiveness of the proposed framework and manifests that introducing suitable weights can boost the performance of recommendation. 
From these figures, $\beta=0.1$ and $\lambda=0.3$ for Sports, $\beta=0.1$ and $\lambda=0.3$ for Beauty, $\beta=0.1$ and $\lambda=0.2$ for Toys, and $\beta=0.1$ and $\lambda=0.1$ for ML-1M are generally proper to ICSRec. 
The weight of $\mathcal{L}_{cicl}$, i.e., $\lambda$, is commonly larger than $\beta$, which demonstrates that CICL gains more importance than FICL.

 \noindent\textbf{Impact of the dropout rate.}
The dropout can be seen as a regularization technique to mitigate model overfitting, and the magnitude of its value is closely related to the data~\cite{dropout_1,dropout_2}. 
From Figure~\ref{fig_b}(c), dropout rate=0.5 for three sparse datasets and dropout rate=0.1 for the ML-1M dataset are generally proper to ICSRec. 
The reason may be that the ML-1M dataset contains longer sequences and the loss of more neurons affects the modeling of user intention. 

\noindent\textbf{Impact of the cluster number $K$.} 
Figure~\ref{fig_b}(d) shows the effect of different numbers of intention prototype clusters on the model. 
This figure shows that the number of users' intentions in different datasets is different, which shows the diversity of users' intentions.
In addition, $K$=256 for Sports, $K$=256 for Beauty, $K$=1024 for Toys, and $K$=512 for ML-1M are generally proper to ICSRec. 

\noindent\textbf{Robustness to Noise Data.}
To verify the robustness of ICSRec against noise interactions, we randomly add a certain proportion(i.e, 5\%, 10\%, 15\%, 20\%) of negative items into the input sequences during testing, and examine the final performance of ICSRec and other baselines. 
From Figure~\ref{noise_exp} we can see that adding noise data deteriorates the performances of four models. 
By comparing SASRec and other models, it can be seen that adding a contrastive auxiliary task can significantly improve the model's robustness. 
By comparing ICSRec and other models, it can be seen that our model performs better than other models. 
Especially, with $10\%$ noise proportion, our model can even outperform other models without noise dataset on four datasets. 
The reason might be that ICSRec leverages contrastive learning to model intentions more accurately from cross subsequences, thus increasing the robustness. 
By looking at the performance of the four models on sparse datasets (sports, beauty, and toys), it can be seen that SASRec and DuoRec are much lower than the intent-based models IOCRec and ICSRec when the proportion of noise data is 20. 
This shows the robustness of the model can be improved by leveraging the user's intention on sparse datasets. 
In addition, comparing the performances of the four models on the dense dataset ml-1m, we can see that 
IOCRec performs less robustly than DuoRec on ML-1M, probably because users' intentions are more diverse in dense datasets, and IOCRec introduces stochastic data augmentation to model the intention, which may destroy the user's original intentions~\cite{MISS}. 
  

\noindent\textbf{Impacts of encoder $f_{\theta}(\cdot)$.} 
To further investigate the effectiveness of our proposed methods, we use other models, such GRU4Rec, as the backbone encoder. 
Specially, we consider the following settings of ICSRec as the backbone encoder for experiments: (B) ICSRec$_{GRU}$: we use the GRU4Rec~\cite{GRU4Rec} as the backbone encoder, (D) ICSRec$_{SAS}$: the default model that uses SASRec~\cite{SASRec} as the backbone encoder. 
Table~\ref{tab:encoder} shows the performance of ICSRec with different encoders, as well as the performance of backbone models. 
It can be seen that (B) and (D) outperform the respective backbone encoder models. 
This indicates that the intent representation learning module is a general module that can help improve the performance of existing SR methods. 
Comparing (A) and (C), (B) and (D), we can find that the encoder dominates the performance of ICSRec, and the intent representation learning module is a complementary part that can help further improve the performance.

\section{Related Work}
\subsection{Sequential Recommendation}
Sequential recommendation (SR) aims to predict successive preferences according to one's historical interactions, which has been heavily researched in academia and industry~\cite{SRS,SRS01}. 
Classical FPM-C~\cite{FPMC} fuses both sequential patterns and users' general interests. 
FOSSIL~\cite{FOSSIL} improves the robustness against sparse data by combining similarity-based methods and high-order Markov Chains. 
With the recent advancements in deep learning, many SR models are gradually being combined with neural networks~\cite{CoCoRec,RNN}. 
Such as Recurrent Neural Networks (RNN) based~\cite{GRU4Rec,RNN} and Convolutional Neural Networks (CNN) based~\cite{Caser}.  
The recent success of Transformer~\cite{Att} advances the development of SR. 
SASRec~\cite{SASRec} first utilizes a transformer to model the users' ordered historical interactions, which has made an immense success.  
LSAN~\cite{LSAN} devises a temporal context-aware embedding and proposes a twine-attention sequential framework. 
STOSA~\cite{STOSA} introduces stochastic embeddings and proposes a Wasserstein Self-Attention in SR. 
However, these models are commonly limited by sparse and noisy problems.

\subsection{Self-supervised Learning for Sequential  Recommendation}
Motivated by the immense success of Self-Supervised Learning (SSL) in the field of Natural Language Process (NLP)~\cite{BERT} and Computer Vision (CV)~\cite{VIT} and its effectiveness in solving data sparsity problems, a growing number of works are now applying SSL to recommendation. 
BERT4Rec~\cite{BERT4Rec} leverages pre-trained BERT to generate the representation of the target item from the user's historical interactions. 
S$^3$-Rec~\cite{S3Rec} utilizes four auxiliary self-supervised tasks to capture the sequential information by maximizing the mutual information of different views. 
More interesting works~\cite{CL4SRec,CoSeRec,DuoRec,MCLRec} introduce Contrastive Learning (CL), which leverages the information from augmented views to boost the effectiveness of learned representations~\cite{SSL}, into SR to alleviate the noise and sparse problems. 
MMInfoRec~\cite{CPC} applies an end-to-end contrastive learning scheme for feature-based SR. 
CL4SRec~\cite{CL4SRec} learns the representations of users by maximizing the agreement between different augmented views of the same user's chronological interactions and optimizing the contrastive loss with the main task simultaneously.  
CoSeRec~\cite{CoSeRec} further improves CL4SRec by introducing two more robust data augmentation methods for generating contrastive pairs. 
More prior works~\cite{GCA,DHCN,SGL} also explore the application of SSL to graph-based recommendation. 
SGL~\cite{SGL} generates two augmented views with graph augmentation and optimizes the node-level contrastive loss.
Different from constructing views by adopting data augmentation, DuoRec~\cite{DuoRec} chooses to construct view pairs with model augmentation, which could maintain the sequential correlations in the process of training. 
However, these augmentation operations are all hand-crafted and may change the original intention behind the sequence. 

\subsection{Intent-guided Recommender Systems}
Many recent approaches have turned their attention to studying users' intentions to improve the performance and robustness of recommender systems~\cite{MIND,MCPRN,IDSR,CoCoRec,ISRec}. 
Some exciting works~\cite{ASLI,CoCoRec,CAFE} construct a complementary task for learning the user's intentions by introducing auxiliary information (e.g., predicting the next item's category, predicting the user's next action type, etc.). 
Since such information may not always be available or truly express the user's intentions, some interesting works have been proposed to model users' intents in the latent space~\cite{ICLRec}.  
DSSRec~\cite{DSSRec} leverages a seq2seq training strategy to extract supervision signals from multiple future interactions and introduces an intent variable to capture mutual information between a user's historical and future behavior sequences. 
SINE~\cite{SINE} proposes a sparse interest extraction module to infer the interacted intentions of a user from a large pool of intention groups.
ICLRec~\cite{ICLRec} extracts users' intent distributions from all user behavior sequences via clustering and integrates the learned intent into the SR model via a contrastive SSL loss. 
IOCRec~\cite{IOCRec} improves ICLRec by introducing two new modules, global and local modules, for modeling intentions, and it leverages these two modules to alleviate the noise problem of the contrastive learning task. 
MITGNN~\cite{MITGNN} proposes a multi-intent translation graph neural network to mine users' multiple intents by considering the correlations of the intents. 
However, these methods all ignore the intent supervisory signals hidden in the user interaction sequence~\cite{GCL4SR}. 
In addition, introducing stochastic data augmentation into intent modeling may change the original intent hidden in the interaction sequence~\cite{MISS}. 
Instead, our method extracts supervisory signals from users' interactions and leverages Contrastive Learning (CL) to learn users' intentions. 
\section{Conclusion}
In this paper, we propose a novel contrastive learning based sequential recommendation system called ICSRec. 
ICSRec extracts coarse-grain intent supervisory signals from all users' historical interaction sequences and then utilizes these intent supervisory signals to construct two auxiliary learning objectives for intent representation learning. 
This approach helps alleviate the sparsity problem of interaction data and presents more suitable items for users with different intentions. 
Finally, the outstanding performance of our model on four real-world datasets demonstrates our mentioned model's superiority over other methods. 
Extensive experiments show that IOCRec achieves state-of-the-art performance against a series
of SOTA solutions. 
For future work, we would like to develop novel auxiliary learning objects for improving the performance of ICSRec. 
Moreover, we are also interested in applying ICSRec to improve the performance of other sequential recommendation models.

\section{Acknowledgments}
This research was partially supported by the NSFC (62376180, 62176175), the major project of natural science research in Universities of Jiangsu Province (21KJA520004), Suzhou Science and Technology Development Program(SYG202328) and the Priority Academic Program Development of Jiangsu Higher Education Institutions.

\bibliographystyle{ACM-Reference-Format}
\bibliography{CLISRec}

\section{Appendix}
\begin{figure}[t]
        \begin{minipage}[t]{0.49\linewidth}
        \centerline{\includegraphics[width=1.0\textwidth]{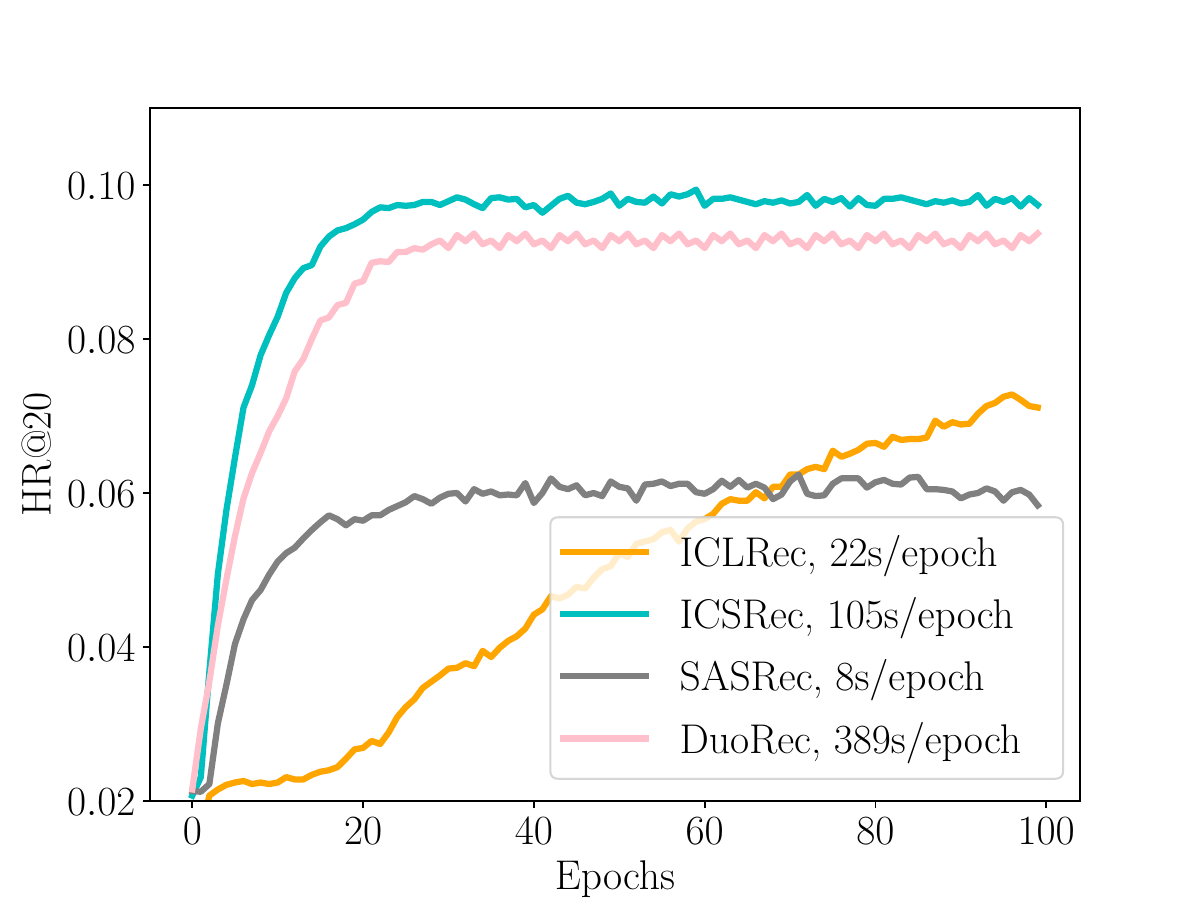}}
     \end{minipage}
        \begin{minipage}[t]{0.49\linewidth}
        \centerline{\includegraphics[width=1.0\textwidth]{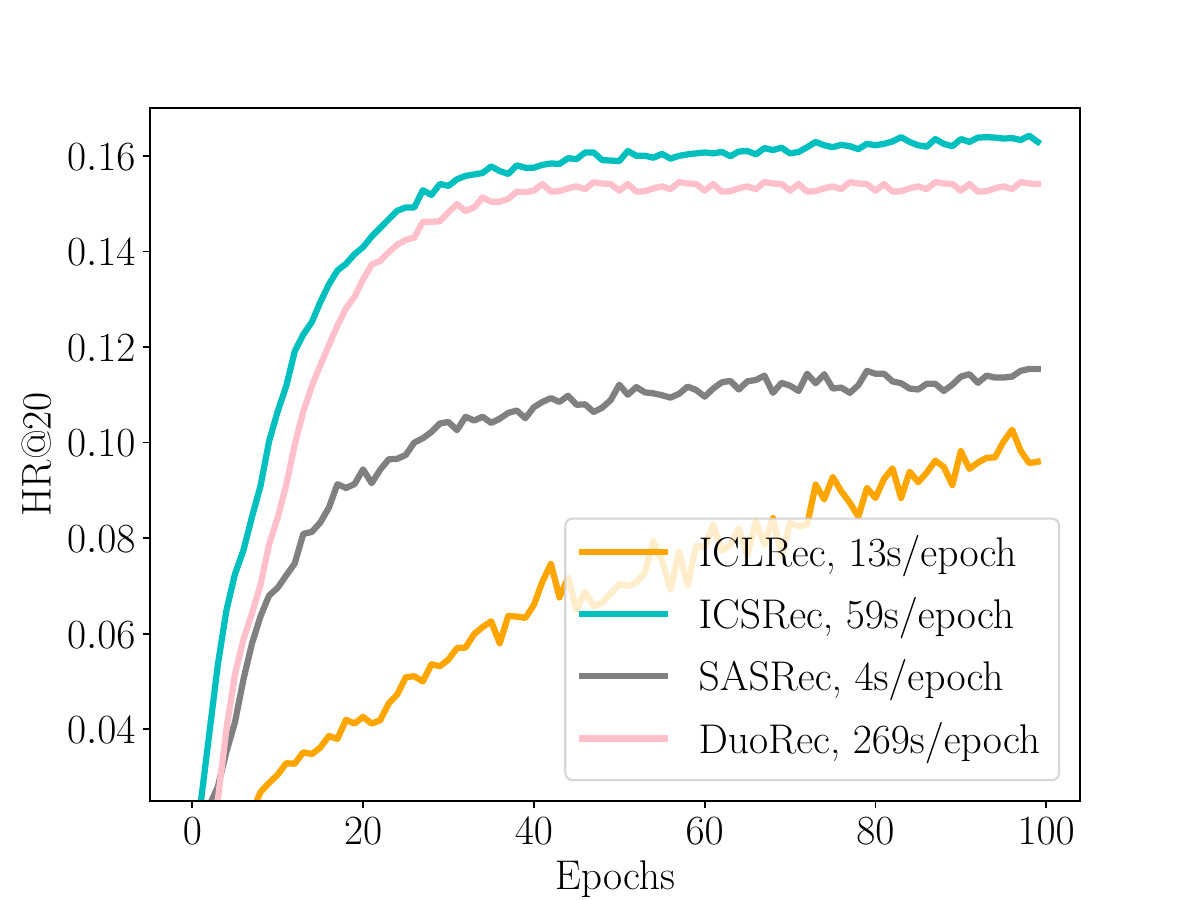}}
     \end{minipage}
\caption{Training efficiency on Sports and Toys datasets.}
\label{epoch_exp}
\end{figure}
\subsection{Case Study}
In this subsection, we will briefly analyze cases that occur in real-world data. 
Specifically, we randomly selected two pairs of users in ML-1M datasets. 
As shown in Table~\ref{tab_intent}, User1 and User2 may have the same intention with the same target item (e.g., watch a drama movie.). 
User3 and User4 have the same target item, but their intentions may be different (e.g., User3 is more inclined to watch the `Action' movie, and User4 is more inclined to watch the `Children' movie.). 
To analyze the effectiveness of our model, we put four sequences into the trained model and put the representation of the four users' output from the model into the visualization.  
The intent heat maps of four users are shown in Figure~\ref{intent}. 
To contrast the heat maps, we use the inner product to calculate the similarity among different intents from the same user. 
Specifically, we use $\mathbf{H}^{u}\cdot(\mathbf{H}^{u})^{T}$ to represent the intent relation among different steps. 
Comparing User1, User2, User3, and User4, we can find that all four users have different intention distributions, which indicates that each user has their own preferences. 
In addition, it can be seen that User1, User3, and User4 all have relatively stable preferences, so their intention representations are more dispersed (i.e., most intentions are related.). 
User2 has more diverse intentions, so his intention representations are more concentrated (i.e., most intentions are unrelated). 
The above findings demonstrate the effectiveness of our model in modeling user intention. 

\subsection{Complexity Analysis} 
In this subsection, we conduct a detailed complexity analysis of ICSRec. We choose SASRec~\cite{SASRec} as our base encoder. Since our ICSRec framework does not introduce any auxiliary learnable parameters. The model size of ICSRec is identical to SASRec. The learned parameters in ICSRec are from embedding and parameters in the self-attention layers, feed-forward networks, and layer normalization. The total number of parameters is $O(|\mathcal{I}|d+nd+d^{2})$.

Then, we analyze the time complexity of the training procedure
of ICSRec framework. In our supervisory signal construction module, we use the dynamic slide window operation to split the original sequence into a set of sub-sequences. We assume the length of each user’s sequence is $L_{a}$ for simplicity. Let $|B|$ denote the size of the mini-batch, and $d$ denote the embedding size. The additional operations mainly
come from three components: applying data split operation
on original user sequences, deriving intent representations from
their interactions in the E-step, and multi-task learning in the M-step.
First, we operate the data by dynamic slide window in the offline. The total complexity is $\mathcal{O}(L_{a}\times(L_{a}-1))|U|$.
Second, for the E-step, the time complicity is $\mathcal{O}(NmKd)$ from clustering, where $N$ represents the number of the processed data after sliding operation, $m$ is the maximum iteration number in clustering($m=20$ in this paper), $K$ is the cluster number and $d$ is the dimensionality of the embedding.
Third, for the M-step, since we have three objectives to optimize the encoder $f_{\theta}(\cdot)$, the time complexity is $3\times \mathcal{O}(N^{2}d+Nd^{2})$. The overall complexity is dominated by the term $\mathcal{O}(3\times (N^{2}d))$, which is 3 times of Transformer based SR with only the next item prediction objective, e.g., SASRec. Fortunately, the model can be benefactive parallized because $f_{\theta}$ is Transformer. In the testing phase, the proposed CICL and FICL objectives are no longer needed, which yields the model to have the same time complexity as SASRec($\mathcal{O}(d|\mathcal{I}|)$). The empirical time spending comparisons are reported in the experiments. 
\begin{algorithm}[t] 
\caption{ The ICSRec Algorithm} 
\label{alg:Framwork} 
\begin{algorithmic}[1] 
\STATE \textbf{Input}: 
training dataset $\{S^{u}\}_{u=1}^{|U|}$, sequence encoder $f_{\theta}(\cdot)$, hyper-parameters $K, \beta, \lambda$, max training epochs $E$, batch size $|B|$.
\STATE \textbf{Output}: $f_{\theta}(\cdot)$.
\STATE Split $\{S^{u}\}_{u=1}^{|U|}$ into several subsequences via Eq.~(\ref{eq2}).
\WHILE{$epoch \leq E$}
\STATE // Update intent prototype representation $C$
\STATE $C=$ Clustering$(f_{\theta}(\{\{S_{t}^{u}\}_{t=1}^{|S^{u}|}\}_{u=1}^{|U|}), K)$
\FOR{a minibatch{$\{s_{u}\}_{u=1}^{|B|}$}}
\FOR{$u \in \{1,2,3,...,|B|\}$}
\STATE // Get sub-sequence $s_2$ with the same target item to $s_1$.
\STATE $s_{2}=Sample(T_{s_1})$.
\STATE // Encoding via $f_{\theta}(\cdot)$.
\STATE $\mathbf{h}^{u}=f_{\theta}(S_{1})$
\STATE $\mathbf{h}_{1}=f_{\theta}(s_{1})$,$\mathbf{h}_{2}=f_{\theta}(s_{2})$
\STATE // Query the intent prototype representations
\STATE $\mathbf{c_{1}}=query(\mathbf{h}_{1})$,$\mathbf{c_{2}}=query(\mathbf{h}_{2})$
\STATE Calculate $\mathcal{L}_{cicl}$ by Eq.~(\ref{eq5})
\STATE Calculate $\mathcal{L}_{ficl}$ by Eq.~(\ref{eq8})
\ENDFOR
\STATE // Optimization the encoder via Eq.~(\ref{eq11})
\STATE $\mathcal{L}=\mathcal{L}_{Rec}+\lambda\cdot \mathcal{L}_{cicl}+\beta\cdot \mathcal{L}_{ficl}$
\STATE Update network $f_{\theta}(\cdot)$ to minimize $\mathcal{L}$
\ENDFOR
\ENDWHILE
\end{algorithmic}
\end{algorithm}

\subsection{Training Efficiency}
Figure~\ref{epoch_exp} demonstrates the efficiency of four compared methods with GPU acceleration. 
On the one hand, SASRec~\cite{SASRec} exhibits the fastest computational speed. 
On the other hand, SASRec~\cite{SASRec} tends to have a less favorable performance compared to other models. 
This indicates that incorporating CL auxiliary tasks not only increases the computational overhead of the model but also enables the model to learn more features, thereby enhancing its recommendation performance.
Upon comparing ICSRec with other models, it becomes evident that the computational overhead of ICSRec is within a reasonable and acceptable range, and its performance improvement is significant, affirming the superiority of ICSRec.

\end{document}